\documentclass[twocolumn, 10pt, aps, superscriptaddress, showpacs, prl, citeautoscript, longbibliography]{revtex4-2}

\usepackage[ngerman, english]{babel}   
\usepackage[utf8]{inputenc}
\usepackage{lmodern}
\usepackage{amssymb}
\usepackage{amsmath}
\usepackage{graphicx}
\usepackage{diagbox} 
\usepackage{hhline}  

\usepackage{xr} 
\usepackage{xcite} 
  
\usepackage[usenames,dvipsnames]{xcolor} 
\usepackage[colorlinks, citecolor={blue}, urlcolor={blue}, linkcolor={blue},hypertexnames=false]{hyperref}

\usepackage{tabularx}                    
\usepackage{multirow, bigdelim}  
\usepackage{longtable}
\usepackage{makecell}                   
\usepackage{cellspace}

\usepackage{mdwlist}      

\usepackage{verbatim}       
\usepackage{bm}
\usepackage{bbm}              
\usepackage{dsfont}			
\usepackage{nicefrac}        
\usepackage[vcentermath]{youngtab} 
\usepackage{cancel}

\usepackage{etoolbox} 

\newcolumntype{L}[1]{>{\raggedright\arraybackslash}p{#1}} 
\newcolumntype{C}[1]{>{\centering\arraybackslash}p{#1}} 
\newcolumntype{R}[1]{>{\raggedleft\arraybackslash}p{#1}} 
\newcommand{\doubleI}{\mathds{1}}

\newcommand{\Sec}[1]{Sec.~\ref{#1}}

\newcommand{\Eq}[1]{Eq.~\eqref{#1}}
\newcommand{\Eqs}[1]{Eqs.~\eqref{#1}}
\newcommand{\Fig}[1]{Fig.~\ref{#1}}

\definecolor{darkgreen}{rgb}{0,0.5,0}
\definecolor{orange}{rgb}{1,0.5,.3}
\definecolor{darkred}{rgb}{.7,0,0}
\definecolor{purple}{rgb}{0.6,0,0.5}
\definecolor{darkpetrol}{RGB}{0,73,76}

\usepackage[normalem]{ulem} 

\newcommand{\setsmalltitle}[1]{{\textit{#1}}---}

\newcommand{\dd}{\mathrm{d}}
\newcommand{\ee}{\mathrm{e}}
\newcommand{\ii}{\mathrm{i}}

     \makeatletter
     \def\maketitle{
     	\@author@finish
     	\title@column\titleblock@produce
     	\suppressfloats[t]}
     \makeatother

\begin{document}

\title{Probing Bardeen--Cooper--Schrieffer Pairing and Quasiparticle Formation in Ultracold Gases by Rydberg Atom Spectroscopy}

\author{Emilio Ramos Rodríguez}
\affiliation{Institut für Theoretische Physik, Universität Heidelberg, 69120 Heidelberg, Germany}
\affiliation{Université Paris Cité, Laboratoire Matériaux et Phénomènes Quantiques (MPQ), CNRS, F-75013, Paris, France}

\author{Marcel Gievers}
\affiliation{Institute of Solid State Physics, TU Wien, 1040 Vienna, Austria}

\author{Richard Schmidt}
\affiliation{Institut für Theoretische Physik, Universität Heidelberg, 69120 Heidelberg, Germany}

\date{\today}
\pacs{}

\begin{abstract}
Locally probing pairing in fermionic superfluids, ranging from micro- to macroscopic scales, has been a long-standing challenge. Here, we investigate a new approach that uses Rydberg impurities as a spectroscopic sensor of the surrounding strongly correlated state of ultracold paired fermions. The extended wave function of the Rydberg electron induces a finite-range potential that can bind atoms from the BCS medium, forming molecular states. As a consequence, the optical absorption spectrum of the impurity encodes key many-body properties. Using the functional determinant approach, we provide a direct measure of the superfluid gap through frequency shifts of dimer and trimer peaks. The spectra also reveal whether the Cooper pairs are broken or trapped intact. For static Rydberg atoms, we relate this signature of pairing to the suppression of the orthogonality catastrophe due to the superconducting gap resulting in the formation of well-defined polaron quasiparticles. Our work establishes Rydberg atom spectroscopy as a powerful local probe of strongly correlated matter.
\end{abstract}

\maketitle

Recent advances in ultracold atom experiments have opened new pathways for exploring quantum many-body phenomena with unprecedented precision and in a controlled setting~\cite{bloch2008many}. Among these, paired two-component fermionic systems, such as described by Bardeen--Cooper--Schrieffer (BCS) theory \cite{PhysRev.108.1175}, remain a cornerstone of condensed matter physics. The central mechanism of the BCS theory, namely the formation of Cooper pairs, underpins the emergence of superconductivity, where fermions undergo pairing to form a collective quantum state. However, despite the theoretical simplicity of BCS theory, existing experimental probes are limited to global, spatially averaged measurements~\cite{PhysRevLett.101.140403,Hoinka2017Goldstone,PhysRevLett.128.100401}, so directly probing the internal structure and dynamics of Cooper pairs remains an experimental challenge.

In solid-state systems, advances in local sensing, such as nitrogen-vacancy centers in diamonds~\cite{Kolkowitz2015Probing,Casola2018Probing}, have enabled access to the properties of material samples from the macroscopic down to the nanometer scale~\cite{PhysRevLett.89.220407}. These methods provide insight into correlated phases by converting many-body effects into measurable quantities of a controlled quantum sensor. 

In ultracold atomic systems, the search continues for local probes that can combine microscopic spatial resolution with the ability to measure spectral features of their quantum environment~\cite{PhysRevX.10.011018,PhysRevB.104.035133,PhysRevLett.129.120404}. Among these, Rydberg atoms have emerged as promising candidates for such local spectroscopic probes. Rydberg atoms, which are characterized by highly excited electronic states and a spatial extent of hundreds of nanometers~\cite{Urban_2009}, induce a finite-range potential in their environment (cf.\ \Fig{fig:Rydberg-BCS}). Through a tunable scattering process, particles from the environment can bind to the Rydberg atom at the Rydberg radius $r_{\mathrm{Ryd}}$, forming ultralong-range Rydberg molecules~\cite{greene2000creation,DeSalvo2015Ultralong,Eiles2019Trilobites,Fey2020Ultralong}. These exotic bound states have been proposed as microscopic probes for the spectral properties of a Bose--Einstein condensate (BEC)~\cite{schlagmuller2016probing}, Fermi gases~\cite{sous2020rydberg} and polaron formation~\cite{Gievers2024Probing}. In addition, for a one-dimensional system in equilibrium, the effect of a Rydberg impurity on the spatial dependence of the BCS gap parameter has been studied~\cite{Chien2024Breaking,PhysRevA.111.043316}.

\begin{figure}
	\centering
	\includegraphics[width=0.48\textwidth]{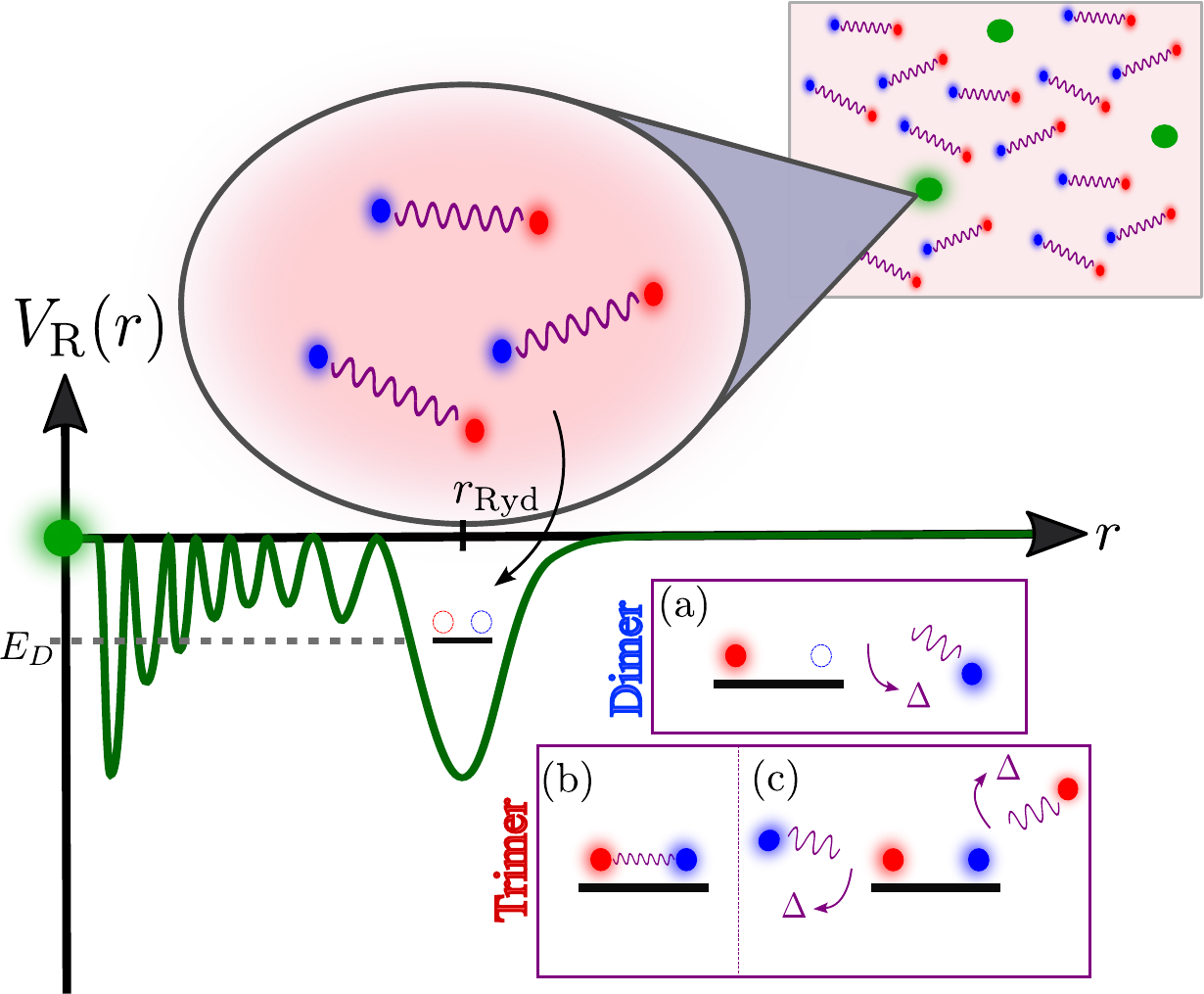}\hfill
	\caption{Illustration of a Rydberg atom immersed in a BCS superfluid. The Rydberg potential $V_\mathrm{R}(\bm{r})$ [\Eq{eq:potential}] is depicted by the green curve and Cooper pairs by red and blue circles connected by a wiggly purple line. A dimer~(a) is formed after breaking a Cooper pair with the energy cost of the gap $\Delta$. A trimer is formed by either~(b) binding a whole Cooper pair or~(c) breaking two Cooper pairs with an energy cost of $2\Delta$.}
	\label{fig:Rydberg-BCS}
\end{figure}
In this Letter, we describe how a heavy Rydberg impurity can be introduced to examine the local properties of paired superfluids (cf.\ \Fig{fig:Rydberg-BCS}). For this, we extend the application of the functional determinant approach~\cite{schmidt2018universal,wang_functional_2023} to anomalous fermion operators to obtain the absorption spectrum of the impurity. We show how each molecular absorption peak is generated by the breaking [cf.\ Figs.~\ref{fig:Rydberg-BCS}(a) and (c)] or trapping [cf.\ \Fig{fig:Rydberg-BCS}(b)] of a Cooper pair. The energy cost of breaking a Cooper pair allows for the measurement of the superconducting gap through shifts in the spectral peaks. 

Because of the large scale separation, Rydberg sensing allows one to probe superfluid dynamics locally resolved and on ultrafast timescales. Strikingly, we find that Rydberg impurities can be used to study the orthogonality catastrophe (OC) and quasiparticle formation in the long-time domain. The OC typically presents itself in heavy-impurity systems, where multiple particle-hole excitations lead to a Fermi edge and power-law decay of the spectrum~\cite{anderson_infrared_1967,Nozieres1969Singularities3,Gievers2025Subleading}. However, a gapped dispersion at the Fermi energy suppresses these excitations, essentially inducing an IR cutoff and enhancing quasiparticle formation~\cite{Chen2025Massgap}. Here, we show how the suppression of the OC allows us to verify intact trapping in the trimer state and quantify the dynamical transition into the quasiparticle regime from short- to long-time scales. With this method, one can combine the long-time dynamics of an impurity in a BCS state~\cite{Wang2022Heavy,Wang2022Exact} with the microscopic probing of individual Cooper pairs in a single experiment.

\setsmalltitle{Model} We consider a three-dimensional Fermi gas with two spin components denoted by $\sigma=\left\{\uparrow,\downarrow\right\}$. The pairing between the components is modeled within the mean-field BCS approximation \cite{PhysRev.108.1175}. The BCS Hamiltonian $\hat{H}_0$, without an impurity, takes the form ($\hbar=1$)
\begin{align}
\hat{H}_0=\int_{\bm{r}}\hat{c}^{\dagger}_{\bm{r}}\begin{pmatrix}
    \hat{h}_\uparrow&\Delta\\
    \Delta &-\hat{h}_\downarrow
\end{pmatrix}\hat{c}_{\bm{r}}+E_g,
\label{eq:hfree}
\end{align}
where $\hat{h}_{\sigma}=-\bm{\nabla}^2/({2m_{\sigma}})-\mu_{\sigma}$, $\mu_\sigma$ is the chemical potential and $m_\sigma$ the mass of the $\sigma$-component. Furthermore, we employ the Nambu spinor representation ${\hat{c}^\dagger_{\bm{r}}=\left(\hat{c}^\dagger_{\uparrow\bm{r}},\, \hat{c}_{\downarrow\bm{r}}\right)}$ where $\hat{c}^\dagger_{\uparrow\bm{r}}$ ($\hat{c}_{\downarrow\bm{r}}$) is the creation (annihilation) operator for a $\uparrow$($\downarrow$) fermion at position $\bm{r}$. The gap $\Delta\in\mathbb{R}^+$ represents the $s$-wave BCS order parameter. Note that $\Delta$ is a single-particle gap parameter in the weak-coupling regime. For strong coupling, however, it refers to the binding energy of tightly bound fermion pairs. The constant $E_g$ arises from the Nambu representation and the mean-field ground state energy (see Supplemental Material~\cite{supplement}).

To introduce the impurity, we consider the following scheme. Initially, the impurity resides in a noninteracting electronic ground state $|0\rangle$, leaving the BCS host system in equilibrium. The impurity is then optically excited to a Rydberg state $|\mathrm{R}\rangle$. Here, the excited electron generates a static potential~\cite{du1987interaction,greene2000creation}:
\begin{equation}
    \label{eq:potential}
    V_{\mathrm{R}}(\bm{r})=\frac{2\pi a_e}{m_e}|\psi_{n_{\mathrm{Ryd}}}(\bm{r})|^2,
\end{equation}
with $a_e$ the electron-neutral atom scattering length, $m_e$ the electron mass, and $\psi_{n_{\mathrm{Ryd}}}(\bm{r})$ the wave function of the electron in a Rydberg state in an $s$-wave shell with principal quantum number $n_{\mathrm{Ryd}}$.

The Hamiltonian for the impurity-BCS system is given by
\begin{equation}
\hat{H}=|\mathrm{0}\rangle\langle\mathrm{0}|\otimes\hat H_{0}+|\mathrm{R}\rangle\langle\mathrm{R}|\otimes\hat{H}_{\mathrm{R}},\label{eq:hamiltonian}
\end{equation}
where
\begin{align}
\hat{H}_{\mathrm{R}}=\int_{\bm{r}}\hat{c}^{\dagger}_{\bm{r}}\begin{pmatrix}
    \hat{h}_{\uparrow,\mathrm{R}}&\Delta\\
    \Delta&-\hat{h}_{\downarrow,\mathrm{R}}
    \end{pmatrix}\hat{c}_{\bm{r}}+\tilde{E}_g\label{eq:hrydberg},
\end{align}
with $\hat{h}_{\sigma,\mathrm{R}}=\hat{h}_{\sigma}+V_{\mathrm{R}}(\bm{r})$. Here, $\hat{H}_{\mathrm{R}}$ describes the BCS state in the presence of the Rydberg potential $V_{\mathrm{R}}(\bm{r})$; $\hat{h}_{\sigma,\mathrm{R}}$ is the corresponding single-particle operator, and $\tilde{E}_g$ accounts for the introduction of \Eq{eq:potential}. 

Following the excitation, the absorption spectrum $A(\omega)$ is measured. Because of its optical nature, the timescale of the excitation can be assumed to be much shorter than the equilibration time between the superfluid and the impurity \cite{bendkowsky2009observation,butscher2010atom,sous2020rydberg}, ensuring that the order parameter $\Delta$ remains homogeneous in the vicinity of the impurity during the procedure. The contribution of inhomogeneities in the BCS superfluid to the sudden absorption signals can, thus, be neglected. We note that recently, in a BCS medium, the backreaction of a Rydberg atom on the spatial dependence of the gap parameter and density has been analyzed for a one-dimensional system~\cite{Chien2024Breaking,PhysRevA.111.043316} whose size scarcely exceeds the Rydberg atom. Here we focus on the experimentally relevant case where the dynamics follows a sudden excitation \cite{cetina2016ultrafast,sous2020rydberg,Gievers2024Probing}, i.e., where the BCS state does not equilibrate with the Rydberg atom. To this end, we consider the out-of-equilibrium dynamics of a three-dimensional configuration with a large system size and compute the actual optical absorption spectra.

To obtain the absorption spectrum, we start by calculating the Ramsey signal~\cite{schmidt2018universal}
\begin{equation}\label{eq:ramsey}
    S(t)=\mathrm{Tr}\left\{\hat{\rho}\, \ee^{\ii\hat{H}_0t}\ee^{-\ii\hat{H}_{\mathrm{R}}t}\right\},
\end{equation}
which quantifies the time-resolved overlap between many-body states with and without the Rydberg potential whose direct measurement has been demonstrated in both neutral~\cite{cetina2016ultrafast} and Rydberg atoms~\cite{manybody,butscher2010atom}. In~\Eq{eq:ramsey}, $\hat{\rho}=\ee^{-\beta\hat{H}_0}/\mathrm{Tr}\left\{\ee^{-\beta\hat{H}_0}\right\}$ is the thermal density matrix and $\beta=1/T$ the inverse temperature. 

From here, the absorption spectrum follows as \cite{schmidt2018universal}
\begin{equation}\label{eq:absorption}
    A(\omega)=2\mathrm{Re}\left\{\int_{0}^{\infty}\!\dd t\,S(t)\,\ee^{\ii\omega t} \right\}.
\end{equation}   
To compute $S(t)$, we employ the functional determinant approach (FDA) \cite{klich2003elementary,schonhammer2007full,schmidt2018universal}, reducing the problem to a single-particle description. We adapt the FDA to the BCS state and numerically solve the Schr\"odinger equation with and without the Rydberg potential by applying the Bogoliubov transformation~\cite{supplement}. While FDA can be directly used for finite temperature, the following results are all given for $T=0$.

\begin{figure*}
    \centering 
    \includegraphics[width=\linewidth]{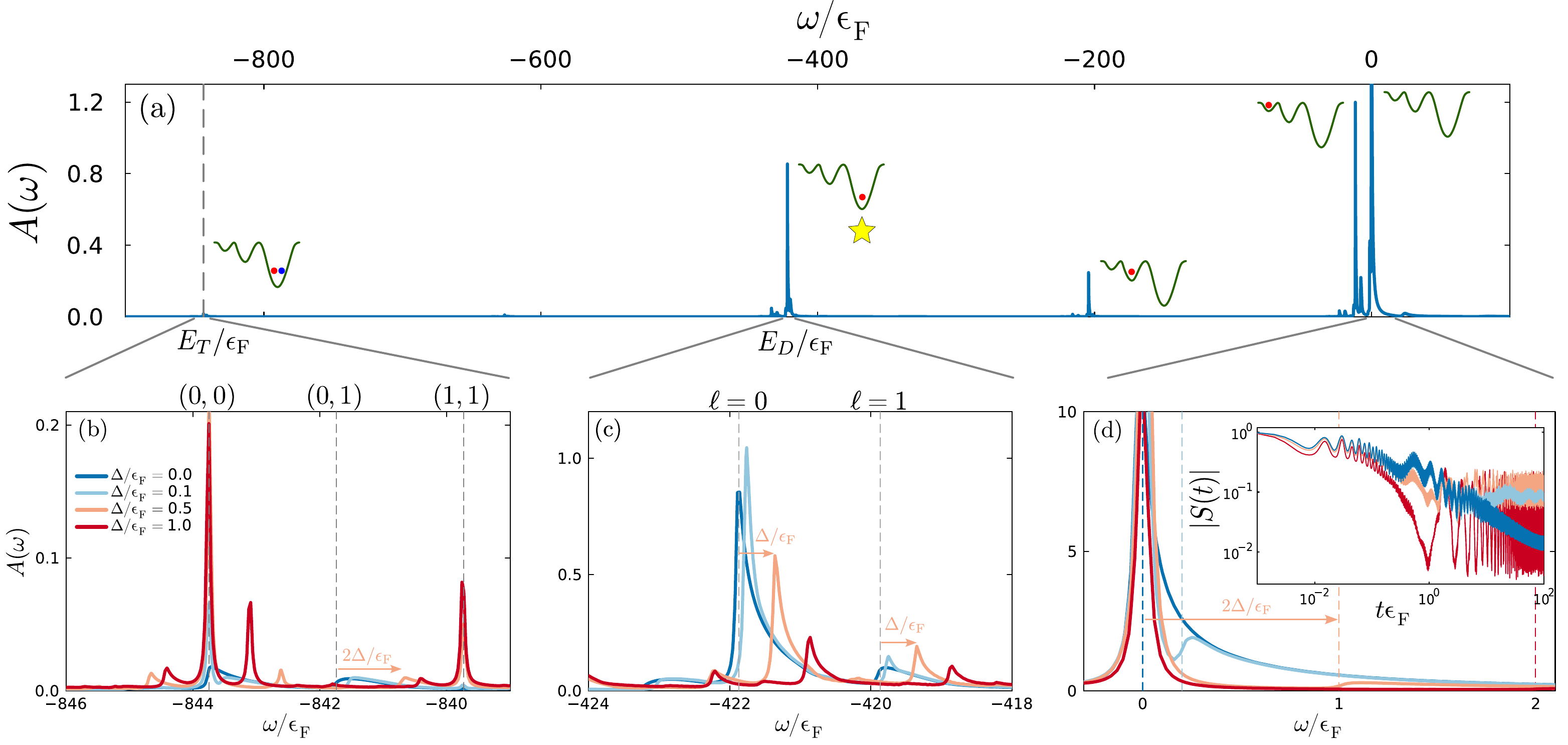}\\
    \vspace{-1em}
    \caption{Rydberg atom spectroscopy. (a) Absorption spectrum for $\Delta=0$, where each peak corresponds to the occupation of a bound state in the Rydberg potential. The energy scale is shifted such that the bare atomic Rydberg peak is at the origin. (b)--(d) study the dependence of the spectrum on the gap parameter $\Delta$ (light blue to red curves). (b) Magnification of the trimer state, i.e., the double occupation of the bound state. Dashed lines mark the energies of different angular-momentum states of the bound state, denoted by $\ell=(\ell_1,\ell_2)$. Peaks with mixed angular momentum feature a shift of $2\Delta$, indicating the rupture of two Cooper pairs. Peaks with the same angular momentum do not show such a shift indicating intact trapping of the Cooper pair. (c) Dimer state: single occupation of the bound state for different angular momenta (dashed lines). All peaks shift proportional to the gap strength $\Delta$, resulting in a direct measure of the superfluid gap, $\Delta=|E^{(\Delta)}_D-E_D|$. (d) Atomic Rydberg peak (repulsive branch): for $\Delta=0$, the orthogonality catastrophe is present as seen from the power-law decay of the spectrum. This feature deforms into a $\delta$-like peak structure for increasing gap strength accompanied by the appearance of a secondary branch corresponding to a Bogoliubov excitation. Inset: Ramsey signal for each gap strength showing the emergence of quasiparticle behavior.}
	\label{fig:spectrum}
\end{figure*}

\setsmalltitle{Rydberg atom spectroscopy} Motivated by experimental setups~\cite{PhysRevA.61.053409,Ryabtsev_2016,PhysRevA.20.507,PhysRevLett.89.053202}, we examine a Rydberg atom of $^{87}$Rb that interacts with a two-component $^{40}$K gas (BCS medium). This setting is experimentally realizable due to the separation of the Feshbach resonances of $^{40}$K-$^{40}$K and $^{40}$K-$^{87}$Rb scattering~\cite{Ospelkaus2006Interaction,Chin2010Feshbach}. For numerical results, the electron of the $^{87}$Rb atom is excited to the $n_{\mathrm{Ryd}}=60$ level and characterized by a scattering length $a_e=-15\,a_0$ \cite{eiles2018formation}, with $a_0$ the Bohr radius. Although our model also generalizes to imbalanced configurations, we consider here the density balanced scenario where $\mu=\mu_{\uparrow}=\mu_{\downarrow}$ and $m=m_{\uparrow}=m_{\downarrow}$ at a background density of $\rho_0=5\times10^{11}\,\mathrm{cm}^{-3}$ with $\epsilon_\mathrm{F}\sim1.2\,\mathrm{kHz}$.

For precision sensing of the BCS state, first a reference spectrum has to be established. This is shown in~\Fig{fig:spectrum}(a) where we consider the absorption spectrum in the absence of interactions, ${\Delta = 0}$, which is in agreement with data from previous research~\cite{sous2020rydberg,Gievers2024Probing}. The general structure is characterized by several absorption peaks at negative frequencies. Each peak corresponds to the occupation of a bound state given by the potential $V_{\mathrm{R}}(\bm{r})$. In our numerical calculations~\cite{supplement}, each scattering channel hosts up to three two-body bound states. The finite-range nature of the potential sets an upper bound for the angular momentum that must be considered; here, $\ell\leq\ell_{\mathrm{max}}=5$. 

Our discussion of the local and time-resolved probing of the BCS state focuses on the outermost bound state, corresponding to the deepest well located at $r_{\mathrm{Ryd}}$ as depicted in~\Fig{fig:Rydberg-BCS} and marked by a yellow star in~\Fig{fig:spectrum}(a). We refer to the single occupation of this bound state as the \textit{dimer} state. For $n=60$, in the $s$-wave channel, it has a binding energy of $E_D\approx-421.9\,\epsilon_\mathrm{F}$; see~\Fig{fig:spectrum}(a). The double occupation of the bound state, the \textit{trimer} state, then has an energy of $E_T=2E_D$; peak at the dashed line in~\Fig{fig:spectrum}(a). For a complete analysis of each peak in the spectrum, including, in particular, those not discussed below, we refer to Supplemental Material~\cite{supplement}.

Let us analyze how the spectrum is modified in the presence of the BCS state, i.e., $\Delta>0$. The dimer peak for different values of the gap $\Delta$ is shown in~\Fig{fig:spectrum}(c). The vertical dashed lines, from left to right, correspond to the angular momentum states of the particle occupying the bound state, $\ell = 0$ and $\ell = 1$, respectively. As the superconducting gap increases, the $\ell = 0$ peak shifts from its value $E_D$ in the absence of $\Delta$ to higher frequencies, denoted as $E_D^{(\Delta)}$. This represents a general feature of the dimer peak; also observed for the $\ell = 1$ state. The single occupation originates from the rupture of a Cooper pair to form the bound state. As the gap increases, so does the binding energy of the Cooper pair. Thus, the impurity requires more energy to trap a single particle of the medium. Indeed, the shift we find is precisely the gap $\Delta$, such that   
\begin{equation}\label{eq:gap}
    \Delta\approx |E^{(\Delta)}_D-E_D|.
\end{equation}
This spectral feature, thus, allows one to measure the gap locally using focused laser light. Note that because the Rydberg atom remains out of equilibrium with the BCS medium, through spatially distributed measurements, it is possible to reconstruct the spatial dependence of the gap parameter $\Delta(x)$.

The trimer state [cf.\ \Fig{fig:spectrum}(b)] offers richer physics. The figure displays three dashed lines, representing the angular momenta of the two particles occupying the bound state, denoted by $(\ell_1,\ell_2)$. Similarly to the dimer state, peaks corresponding to states with mixed angular momenta, e.g., (0,1), exhibit a shift of $2\Delta$. Strikingly, peaks associated with states where both particles share the same angular momentum, e.g., $(0,0)$ and $(1,1)$, do not shift as the superfluid gap $\Delta$ increases.

In the first case, the bounded particles with different angular momenta originate from the rupture of two Cooper pairs, as reflected by the $2\Delta$ shift. However, an intact Cooper pair can also bind to the impurity as a whole. In this case, the trapping of the Cooper pair does not require an energy cost beyond the Rydberg molecular energy. As a result, this process is independent of the binding energy of the Cooper pair and will not lead to a shift in the spectrum, as seen in~\Fig{fig:spectrum}(b). Remarkably, we find that Cooper-pair trapping occurs even when its coherence length, ${\xi(\Delta/\epsilon_\mathrm{F})\approx \hbar v_\mathrm{F}/\Delta}$, is greater than the diameter of the Rydberg potential, e.g., ${\xi(\Delta/\epsilon_\mathrm{F}=0.1)\sim2.06\,\mu\mathrm{m} >2r_{\mathrm{Ryd}}\sim0.711\,\mu\mathrm{m}}$. Here, $\xi$ represents only the scale for the spatial decay of the Cooper-pair wave function. We attribute this observation to the three-dimensional geometry. In contrast to the one-dimensional analysis in Refs.~\cite{Chien2024Breaking,PhysRevA.111.043316}, Cooper pairs can be aligned in multiple ways around the spherical surface defined by the Rydberg radius $r_\mathrm{Ryd}$. Our results thus demonstrate that Cooper pairs can be trapped without disrupting its coherence. Moreover, Figs.~\ref{fig:spectrum}(b)--(c) suggest that the spectral weight of the trimer peak increases with the gap strength while the spectral weight of the dimer peak decreases. Accordingly, it is more probable that Cooper-pair trapping occurs when $\xi$ is smaller, whereas for larger $\xi$, Cooper-pair breaking is preferred. In order to further investigate this mechanism, one could study how the wave function of the Cooper pair changes in response to the static Rydberg potential. This would require variational wave-function-based techniques, which is left for future research. 

As seen in Figs.~\ref{fig:spectrum}(b) and (c), intact trapping or breaking of a Cooper pair leads to distinct spectral profiles, either to a quasiparticle peak or to a power-law tail. The dimer state arises by breaking a Cooper pair, thus creating a Bogoliubov particle-hole excitation across the superconducting gap. Analogous to excitations in a semiconductor, the possible configurations of this particle-hole excitation in the dispersion lead to the characteristic tail seen in the dimer spectrum. In contrast, in the case of intact Cooper-pair trapping, additional particle-hole excitations are suppressed by the BCS gap, leading to the emergence of a quasiparticle peak.

\begin{figure}
	\centering
	\includegraphics[width=0.48\textwidth]{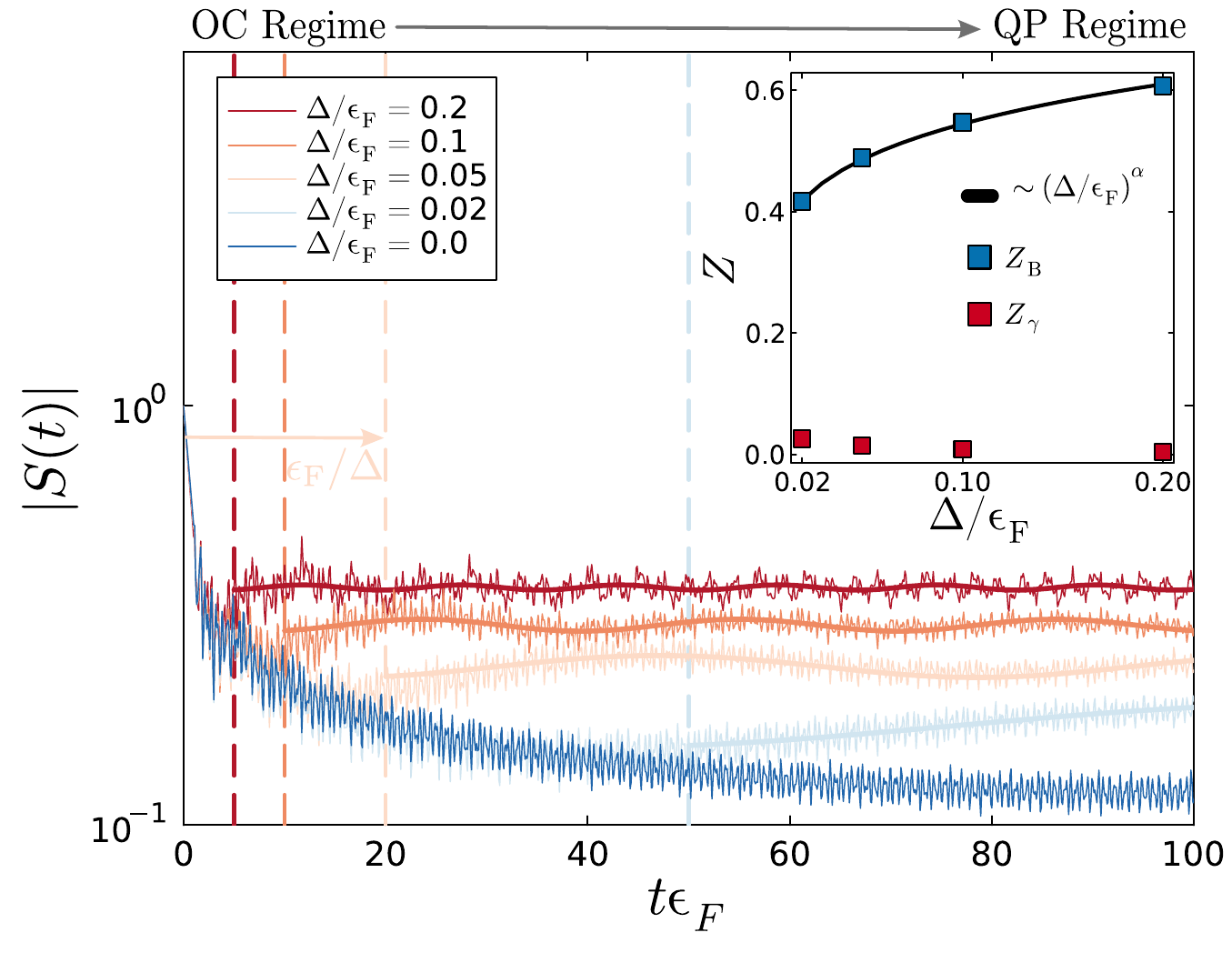}\hfill
	\caption{Orthogonality catastrophe. Ramsey signal for different gap strengths showing the OC for $\Delta=0$ and the emergence of a quasiparticle weight for $\Delta>0$ where the signal is fitted (thick lines) to~\Eq{eq:fit} at times $t\gtrsim1/\Delta$ (dashed lines). Inset: quasiparticle weights of the bath and Bogoliubov excitation, where $Z_B$ follows a power-law relation with scaling given by the scattering shift exponent [\Eq{eq:exp}].}
	\label{fig:oc_full}
\end{figure}
\setsmalltitle{Orthogonality catastrophe (OC)} We now show that Rydberg impurities in a BCS superfluid allow one to study the stabilization of quasiparticle behavior by emerging many-body energy gaps. In \Fig{fig:spectrum}(d), we show the spectrum of the excited branch, where no bound state is occupied, for different gap strengths; we refer to this branch as the atomic Rydberg peak. At $\Delta=0$, the characteristic power-law decay of an immobile impurity can be observed, which is not affected by the microscopic details of the scattering potential~\cite{anderson_infrared_1967,Nozieres1969Singularities3}. However, as the gap increases, the spectrum loses this attribute. Instead, the spectrum deforms from
a power law into a broadened $\delta$-like quasiparticle peak
structure which we attribute to the gapped dispersion of the BCS bath suppressing low-energy particle-hole excitations. The gapped dispersion allows for a clear distinction of the physical mechanisms associated with the trimer and dimer peaks. When trapping an intact Cooper pair, low-energy excitations in the bath are suppressed by the energy gap, which results in the sharp trimer peak. On the other hand, disrupting a Cooper pair leads to Bogoliubov excitations, which characterize the tail of the dimer peak. 
Indeed, the superfluid gap plays an analogous role to a mass gap generated by impurity recoil in the case of a mobile impurity in a Fermi sea~\cite{Chen2025Massgap}. In the Ramsey signal [inset of~\Fig{fig:spectrum}(d)] this is captured by the nondecaying signals for $\Delta>0$.

As the gap increases, the appearance of a secondary branch becomes visible. This branch is the result of the excitation of a Bogoliubov mode from the lower branch of the Bogoliubov dispersion into the upper branch. The excitation must overcome the separation between the branches of the dispersion, reflected by the $2\Delta$ shift. As the gap increases, the excitation becomes energetically less favorable, and the upper branch loses spectral weight, which is transferred to the lower quasiparticle branch.

The gap parameter dictates the timescale on which the Ramsey signal starts to deviate from the power-law decay, or OC regime, and transitions into a dynamical regime associated with quasiparticle formation. This is analyzed in~\Fig{fig:oc_full}, where dashed lines show the timescales given by $\sim1/\Delta$. In order to quantitatively study the impact of the gapped dispersion in the suppression of the OC and quasiparticle formation, we fit the Ramsey signal to the asymptotic form (solid lines in~\Fig{fig:oc_full}):   
\begin{align}
    S(t\, \gtrsim\,1/\Delta)\approx Z_B\ee^{\ii\omega_B t}+Z_\gamma \ee^{\ii\omega_\gamma t}\ .
    \label{eq:fit}
\end{align}
Here, $Z_B$ and $Z_\gamma$ represent the quasiparticle weights of the atomic and Bogoliubov excitation peaks, respectively, with $2\Delta\approx|\omega_B-\omega_\gamma|$. As can be seen from~\Fig{fig:oc_full}, this fit describes the numerical data remarkably well. We find a power-law dependence of the quasiparticle weight of the bath with respect to the gap, leading to $Z_B^{(\Delta)}\sim(\Delta/\epsilon_{\mathrm{F}})^\alpha$ (see the inset in \Fig{fig:oc_full}). This suppression of the orthogonality catastrophe has also been identified previously in a polaron-BCS system \cite{Wang2022Heavy,Wang2022Exact}. Here, we find that the scaling exponent is precisely given by
\begin{equation}
    \label{eq:exp}    \alpha=\sum_\ell(2\ell+1)\left[\frac{\delta_\ell(\epsilon_{\mathrm{F}})}{\pi}\right]^2,
\end{equation}
where $\delta_\ell(\epsilon_{\mathrm{F}})$ is the scattering phase shift of the $\ell$-scattering channel evaluated at the Fermi energy. Therefore, we recover the universal low-energy features of the OC \cite{Wang2022Heavy,Wang2022Exact} while simultaneously resolving the underlying dynamics of Cooper pairing at the microscopic level.

\setsmalltitle{Conclusion}
In this work, we demonstrated that Rydberg impurities can serve as a new comprehensive sensor to locally probe BCS superfluids. We have shown how the absorption spectrum provides critical insights into Cooper pairing, introducing an experimentally realizable tool to directly measure the gap parameter. Our real-time analysis further suggests that intact Cooper-pair trapping manifests as gap-independent spectral features with suppression of the orthogonality catastrophe and the emergence of a well-defined quasiparticle. The approach can be directly extended to study the dependence of the gap parameter on finite temperature and imbalanced mixtures. Future studies can extend this framework to explore the dynamics of a mobile impurity~\cite{Chen2025Massgap} and other strongly correlated systems. Moreover, to obtain more precise quantitative results in the strong-coupling regime, our mean-field BCS analysis may be extended by a self-consistent diagrammatic approach.

\vspace{2mm}
\setsmalltitle{Acknowledgments}
We thank Márton Kanász-Nagy for fruitful discussions. We acknowledge funding for E.R.R.\ and R.S.\ by the DFG (German Research Foundation) – Project-ID 273811115 – SFB 1225 ISOQUANT, the DFG Priority Programme of Giant Interactions in Rydberg Systems (GiRyd) and under Germany's Excellence Strategy EXC 2181/1 - 390900948 (the Heidelberg STRUCTURES Excellence Cluster). The
authors acknowledge TU Wien Bibliothek for financial support through its Open Access Funding
Programme.

\vspace{2mm}
\setsmalltitle{Data availibility}
The data that support the findings of this article are not publicly available. The data are available from the authors upon reasonable request.

\bibliographystyle{apsrev4-2}
\bibliography{main}

@article{PhysRevLett.101.140403,
  title = {Determination of the Superfluid Gap in Atomic Fermi Gases by Quasiparticle Spectroscopy},
  author = {Schirotzek, Andr\'e and Shin, Yong Il and Schunck, Christian H. and Ketterle, Wolfgang},
  journal = {Phys. Rev. Lett.},
  volume = {101},
  issue = {14},
  pages = {140403},
  numpages = {4},
  year = {2008},
  month = {Oct},
  publisher = {American Physical Society},
  doi = {10.1103/PhysRevLett.101.140403},
  url = {https://link.aps.org/doi/10.1103/PhysRevLett.101.140403}
}

@article{benjamin2013quasiparticletheoryresonantinelastic,
  title = {Single-Band Model of Resonant Inelastic X-Ray Scattering by Quasiparticles in High-${T}_{c}$ Cuprate Superconductors},
  author = {Benjamin, David and Klich, Israel and Demler, Eugene},
  journal = {Phys. Rev. Lett.},
  volume = {112},
  issue = {24},
  pages = {247002},
  numpages = {5},
  year = {2014},
  month = {Jun},
  publisher = {American Physical Society},
  doi = {10.1103/PhysRevLett.112.247002},
  url = {https://link.aps.org/doi/10.1103/PhysRevLett.112.247002}
}

@article{Urban_2009,
   title={Observation of Rydberg blockade between two atoms},
   volume={5},
   ISSN={1745-2481},
   url={http://dx.doi.org/10.1038/nphys1178},
   DOI={10.1038/nphys1178},
   number={2},
   journal={Nature Physics},
   publisher={Springer Science and Business Media LLC},
   author={Urban, E. and Johnson, T. A. and Henage, T. and Isenhower, L. and Yavuz, D. D. and Walker, T. G. and Saffman, M.},
   year={2009},
   month=jan, pages={110–114} }

@book{griffiths_introduction_2018,
	address = {Cambridge, England},
	edition = {Third edition},
	title = {Introduction to quantum mechanics},
	isbn = {978-1-107-18963-8},
	publisher = {Cambridge University Press},
	author = {Griffiths, David J. and Schroeter, Darrell F.},
	year = {2018},
}

@article{Combescot1971Infrared,
  title = {Infrared Catastrophe and Excitons in the {X}-ray Spectra of Metals},
  author = {Combescot, M. and Nozi{\`e}res, P.},
  journal = {Journal de Physique},
  volume = {32},
  number = {11-12},
  pages = {913--929},
  year = {1971},
  doi = {10.1051/jphys:019710032011-12091300},
  url = {https://doi.org/10.1051/jphys:019710032011-12091300}
}

@article{manybody,
	author = {Zeiher, Johannes and van Bijnen, Rick and Schau{\ss}, Peter and Hild, Sebastian and Choi, Jae-yoon and Pohl, Thomas and Bloch, Immanuel and Gross, Christian},
	date = {2016/12/01},
	doi = {10.1038/nphys3835},
	id = {Zeiher2016},
	isbn = {1745-2481},
	journal = {Nat.~Phys.},
	number = {12},
	pages = {1095--1099},
	title = {Many-body interferometry of a Rydberg-dressed spin lattice},
	url = {https://doi.org/10.1038/nphys3835},
	volume = {12},
	year = {2016}
}

@article{PhysRevB.104.035133,
  title = {Tracing non-Abelian anyons via impurity particles},
  author = {Baldelli, Niccol\`o and Juli\'a-D\'{\i}az, Bruno and Bhattacharya, Utso and Lewenstein, Maciej and Gra\ss{}, Tobias},
  journal = {Phys. Rev. B},
  volume = {104},
  issue = {3},
  pages = {035133},
  numpages = {10},
  year = {2021},
  month = {Jul},
  publisher = {American Physical Society},
  doi = {10.1103/PhysRevB.104.035133},
  url = {https://link.aps.org/doi/10.1103/PhysRevB.104.035133}
}

@article{PhysRevX.10.011018,
  title = {Single-Atom Quantum Probes for Ultracold Gases Boosted by Nonequilibrium Spin Dynamics},
  author = {Bouton, Quentin and Nettersheim, Jens and Adam, Daniel and Schmidt, Felix and Mayer, Daniel and Lausch, Tobias and Tiemann, Eberhard and Widera, Artur},
  journal = {Phys. Rev. X},
  volume = {10},
  issue = {1},
  pages = {011018},
  numpages = {13},
  year = {2020},
  month = {Jan},
  publisher = {American Physical Society},
  doi = {10.1103/PhysRevX.10.011018},
  url = {https://link.aps.org/doi/10.1103/PhysRevX.10.011018}
}

@article{PhysRevLett.129.120404,
  title = {Coherent and Dephasing Spectroscopy for Single-Impurity Probing of an Ultracold Bath},
  author = {Adam, Daniel and Bouton, Quentin and Nettersheim, Jens and Burgardt, Sabrina and Widera, Artur},
  journal = {Phys. Rev. Lett.},
  volume = {129},
  issue = {12},
  pages = {120404},
  numpages = {6},
  year = {2022},
  month = {Sep},
  publisher = {American Physical Society},
  doi = {10.1103/PhysRevLett.129.120404},
  url = {https://link.aps.org/doi/10.1103/PhysRevLett.129.120404}
}

@article{PhysRevLett.89.220407,
  title = {High-Temperature Superfluidity of Fermionic Atoms in Optical Lattices},
  author = {Hofstetter, W. and Cirac, J. I. and Zoller, P. and Demler, E. and Lukin, M. D.},
  journal = {Phys. Rev. Lett.},
  volume = {89},
  issue = {22},
  pages = {220407},
  numpages = {4},
  year = {2002},
  month = {Nov},
  publisher = {American Physical Society},
  doi = {10.1103/PhysRevLett.89.220407},
  url = {https://link.aps.org/doi/10.1103/PhysRevLett.89.220407}
}

@article{PhysRevLett.89.053202,
  title = {Collisional Properties of Ultracold K-Rb Mixtures},
  author = {Ferrari, G. and Inguscio, M. and Jastrzebski, W. and Modugno, G. and Roati, G. and Simoni, A.},
  journal = {Phys. Rev. Lett.},
  volume = {89},
  issue = {5},
  pages = {053202},
  numpages = {4},
  year = {2002},
  month = {Jul},
  publisher = {American Physical Society},
  doi = {10.1103/PhysRevLett.89.053202},
  url = {https://link.aps.org/doi/10.1103/PhysRevLett.89.053202}
}

@article{PhysRevA.20.507,
  title = {Experimental studies of high-lying Rydberg states in atomic rubidium},
  author = {Liberman, S. and Pinard, J.},
  journal = {Phys. Rev. A},
  volume = {20},
  issue = {2},
  pages = {507--518},
  numpages = {0},
  year = {1979},
  month = {Aug},
  publisher = {American Physical Society},
  doi = {10.1103/PhysRevA.20.507},
  url = {https://link.aps.org/doi/10.1103/PhysRevA.20.507}
}

@article{Ryabtsev_2016,
doi = {10.3367/UFNe.0186.201602k.0206},
url = {https://doi.org/10.3367/UFNe.0186.201602k.0206},
year = {2016},
month = {feb},
publisher = {Turpion Ltd and the Russian Academy of Sciences},
volume = {59},
number = {2},
pages = {196},
author = {Ryabtsev, I I and Beterov, I I and Tretyakov, D B and Entin, V M and Yakshina, E A},
title = {Spectroscopy of cold rubidium Rydberg atoms for applications in quantum information},
journal = {Phys.-Usp.}
}

@article{PhysRevA.61.053409,
  title = {Cooper pairing in ultracold ${}^{40}\mathrm{K}$ using Feshbach resonances},
  author = {Bohn, John L.},
  journal = {Phys. Rev. A},
  volume = {61},
  issue = {5},
  pages = {053409},
  numpages = {4},
  year = {2000},
  month = {Apr},
  publisher = {American Physical Society},
  doi = {10.1103/PhysRevA.61.053409},
  url = {https://link.aps.org/doi/10.1103/PhysRevA.61.053409}
}

@article{PhysRev.108.1175,
  title = {Theory of Superconductivity},
  author = {Bardeen, J. and Cooper, L. N. and Schrieffer, J. R.},
  journal = {Phys. Rev.},
  volume = {108},
  issue = {5},
  pages = {1175--1204},
  numpages = {0},
  year = {1957},
  month = {Dec},
  publisher = {American Physical Society},
  doi = {10.1103/PhysRev.108.1175},
  url = {https://link.aps.org/doi/10.1103/PhysRev.108.1175}
}

@article{sous2020rydberg,
	title={Rydberg impurity in a Fermi gas: Quantum statistics and rotational blockade},
	author={Sous, John and Sadeghpour, H. R. and Killian, T. C. and Demler, Eugene and Schmidt, Richard},
	journal={Phys. Rev. Res.},
	volume={2},
	number={2},
	pages={023021},
	year={2020},
	publisher={APS},
	url={https://journals.aps.org/prresearch/abstract/10.1103/PhysRevResearch.2.023021}
}

@article{schmidt2018universal,
	title={Universal many-body response of heavy impurities coupled to a Fermi sea: a review of recent progress},
	author={Schmidt, Richard and Knap, Michael and Ivanov, Dmitri A and You, Jhih-Shih and Cetina, Marko and Demler, Eugene},
	journal={Reports on Progress in Physics},
	volume={81},
	number={2},
	pages={024401},
	year={2018},
	publisher={IOP Publishing},
	url = {https://iopscience.iop.org/article/10.1088/1361-6633/aa9593/meta}
}

@article{anderson_infrared_1967,
	title = {Infrared Catastrophe in {Fermi} Gases with Local Scattering Potentials},
	volume = {18},
	copyright = {http://link.aps.org/licenses/aps-default-license},
	issn = {0031-9007},
	url = {https://link.aps.org/doi/10.1103/PhysRevLett.18.1049},
	doi = {10.1103/PhysRevLett.18.1049},
	number = {24},
	urldate = {2024-10-27},
	journal = {Phys. Rev. Lett.},
	author = {Anderson, P. W.},
	month = jun,
	year = {1967},
	pages = {1049--1051},
}

@article{Gievers2025Subleading,
  title = {Subleading logarithmic behavior in the parquet formalism},
  author = {Gievers, Marcel and Schmidt, Richard and von Delft, Jan and Kugler, Fabian B.},
  journal = {Phys. Rev. B},
  volume = {111},
  issue = {8},
  pages = {085151},
  numpages = {27},
  year = {2025},
  month = {Feb},
  publisher = {American Physical Society},
  doi = {10.1103/PhysRevB.111.085151},
  url = {https://link.aps.org/doi/10.1103/PhysRevB.111.085151}
}

@article{PhysRevA.111.043316,
  title = {Tunable pairing with local spin-dependent Rydberg molecule potentials in an atomic Fermi superfluid},
  author = {Chien, Chih-Chun and Rittenhouse, Seth T. and Mistakidis, S. I. and Sadeghpour, H. R.},
  journal = {Phys. Rev. A},
  volume = {111},
  issue = {4},
  pages = {043316},
  numpages = {13},
  year = {2025},
  month = {Apr},
  publisher = {American Physical Society},
  doi = {10.1103/PhysRevA.111.043316},
  url = {https://link.aps.org/doi/10.1103/PhysRevA.111.043316}
}

@article{wang_functional_2023,
	title = {Functional determinant approach investigations of heavy impurity physics},
	volume = {33},
	issn = {2309-4710},
	url = {https://link.springer.com/10.1007/s43673-023-00092-5},
	doi = {10.1007/s43673-023-00092-5},
	
	number = {1},
	urldate = {2024-10-27},
	journal = {AAPPS Bull.},
	author = {Wang, J.},
	month = sep,
	year = {2023},
	pages = {20},
}

@article{Nozieres1969Singularities3,
	title = {Singularities in the {X}-ray absorption and emission of metals. {III}. {One}-body theory exact solution},
	volume = {178},
	url = {https://journals.aps.org/pr/abstract/10.1103/PhysRev.178.1097},
	number = {3},
	journal = {Physical Review},
	author = {Nozières, P. and Dominicis, C. T. de},
	year = {1969},
	pages = {1097},
}

@article{cetina2016ultrafast,
  title={Ultrafast many-body interferometry of impurities coupled to a Fermi sea},
  author={Cetina, Marko and Jag, Michael and Lous, Rianne S and Fritsche, Isabella and Walraven, Jook TM and Grimm, Rudolf and Levinsen, Jesper and Parish, Meera M and Schmidt, Richard and Knap, Michael and others},
  journal={Science},
  volume={354},
  number={6308},
  pages={96--99},
  year={2016},
  publisher={American Association for the Advancement of Science},
  url={https://www.science.org/doi/abs/10.1126/science.aaf5134}
}

@article{eiles2018formation,
  title={Formation of long-range Rydberg molecules in two-component ultracold gases},
  author={Eiles, Matthew T},
  journal={Phys. Rev. A},
  volume={98},
  number={4},
  pages={042706},
  year={2018},
  publisher={APS},
  url={https://journals.aps.org/pra/abstract/10.1103/PhysRevA.98.042706}
}

@Inbook{klich2003elementary,
    author="Klich, I.",
    editor="Nazarov, Yuli V.",
    title="An Elementary Derivation of {L}evitov's Formula",
    bookTitle="Quantum Noise in Mesoscopic Physics",
    year="2003",
    publisher="Springer Netherlands",
    address="Dordrecht",
    pages="397--402",
    isbn="978-94-010-0089-5",
    doi="10.1007/978-94-010-0089-5_19",
    url="https://doi.org/10.1007/978-94-010-0089-5_19"
}

@article{schonhammer2007full,
  title={Full counting statistics for noninteracting fermions: Exact results and the Levitov-Lesovik formula},
  author={Sch{\"o}nhammer, Kurt},
  journal={Phys. Rev. B},
  volume={75},
  number={20},
  pages={205329},
  year={2007},
  publisher={APS},
  url={https://journals.aps.org/prb/abstract/10.1103/PhysRevB.75.205329}
}

@article{schlagmuller2016probing,
  title={Probing an electron scattering resonance using Rydberg molecules within a dense and ultracold gas},
  author={Schlagm{\"u}ller, Michael and Liebisch, Tara Cubel and Nguyen, Huan and Lochead, Graham and Engel, Felix and B{\"o}ttcher, Fabian and Westphal, Karl M and Kleinbach, Kathrin S and L{\"o}w, Robert and Hofferberth, Sebastian and others},
  journal={Phys. Rev. Lett.},
  volume={116},
  number={5},
  pages={053001},
  year={2016},
  publisher={APS},
  url={https://journals.aps.org/prl/abstract/10.1103/PhysRevLett.116.053001}
}

@article{greene2000creation,
  title={Creation of polar and nonpolar ultra-long-range Rydberg molecules},
  author={Greene, Chris H and Dickinson, A. S and Sadeghpour, H. R},
  journal={Phys. Rev. Lett.},
  volume={85},
  number={12},
  pages={2458},
  year={2000},
  publisher={APS},
  url={https://journals.aps.org/prl/abstract/10.1103/PhysRevLett.85.2458}
}

@article{du1987interaction,
  title={Interaction between a Rydberg atom and neutral perturbers},
  author={Du, Ning Yi and Greene, Chris H},
  journal={Phys. Rev. A},
  volume={36},
  number={2},
  pages={971},
  year={1987},
  publisher={APS},
  url={https://journals.aps.org/pra/abstract/10.1103/PhysRevA.36.971}
}

@BIBNOTE{supplement,
  note = {See Supplemental Material for additional details; Supplemental Material includes Refs.~\cite{greene2000creation,Gievers2024Probing,anderson_infrared_1967,Chen2025Massgap,Wang2022Heavy,Wang2022Exact,klich2003elementary,benjamin2013quasiparticletheoryresonantinelastic,Combescot1971Infrared,griffiths_introduction_2018}}
}

@article{butscher2010atom,
  title={Atom--molecule coherence for ultralong-range Rydberg dimers},
  url = {https://www.nature.com/articles/nphys1828},
  author={Butscher, Bj{\"o}rn and Nipper, Johannes and Balewski, Jonathan B and Kukota, Ludmila and Bendkowsky, Vera and L{\"o}w, Robert and Pfau, Tilman},
  journal={Nature Physics},
  volume={6},
  number={12},
  pages={970--974},
  year={2010},
  publisher={Nature Publishing Group UK London}
}

@article{bendkowsky2009observation,
  title={Observation of ultralong-range Rydberg molecules},
  url = {https://www.nature.com/articles/nature07945},
  author={Bendkowsky, Vera and Butscher, Bj{\"o}rn and Nipper, Johannes and Shaffer, James P and L{\"o}w, Robert and Pfau, Tilman},
  journal={Nature},
  volume={458},
  number={7241},
  pages={1005--1008},
  year={2009},
  publisher={Nature Publishing Group UK London}
}

@article{Wang2022Heavy,
  title = {Heavy polarons in ultracold atomic Fermi superfluids at the BEC-BCS crossover: Formalism and applications},
  author = {Wang, Jia and Liu, Xia-Ji and Hu, Hui},
  journal = {Phys. Rev. A},
  volume = {105},
  issue = {4},
  pages = {043320},
  numpages = {13},
  year = {2022},
  month = {Apr},
  publisher = {American Physical Society},
  doi = {10.1103/PhysRevA.105.043320},
  url = {https://link.aps.org/doi/10.1103/PhysRevA.105.043320}
}

@article{bloch2008many,
  title={Many-body physics with ultracold gases},
  author={Bloch, Immanuel and Dalibard, Jean and Zwerger, Wilhelm},
  journal={Rev. Mod. Phys.},
  volume={80},
  number={3},
  pages={885},
  year={2008},
  publisher={APS},
  url={https://journals.aps.org/rmp/abstract/10.1103/RevModPhys.80.885}
}

@article{Gievers2024Probing,
  title = {Probing Polaron Clouds by Rydberg Atom Spectroscopy},
  author = {Gievers, Marcel and Wagner, Marcel and Schmidt, Richard},
  journal = {Phys. Rev. Lett.},
  volume = {132},
  issue = {5},
  pages = {053401},
  numpages = {6},
  year = {2024},
  month = {Jan},
  publisher = {American Physical Society},
  doi = {10.1103/PhysRevLett.132.053401},
  url = {https://link.aps.org/doi/10.1103/PhysRevLett.132.053401}
}

@article{Chien2024Breaking,
  title = {Breaking and trapping Cooper pairs by Rydberg-molecule spectroscopy in atomic Fermi superfluids},
  author = {Chien, Chih-Chun and Mistakidis, S. I. and Sadeghpour, H. R.},
  journal = {Phys. Rev. A},
  volume = {110},
  issue = {5},
  pages = {L051303},
  numpages = {7},
  year = {2024},
  month = {Nov},
  publisher = {American Physical Society},
  doi = {10.1103/PhysRevA.110.L051303},
  url = {https://link.aps.org/doi/10.1103/PhysRevA.110.L051303}
}

@article{Wang2022Exact,
  title = {Exact Quasiparticle Properties of a Heavy Polaron in BCS Fermi Superfluids},
  author = {Wang, Jia and Liu, Xia-Ji and Hu, Hui},
  journal = {Phys. Rev. Lett.},
  volume = {128},
  issue = {17},
  pages = {175301},
  numpages = {7},
  year = {2022},
  month = {Apr},
  publisher = {American Physical Society},
  doi = {10.1103/PhysRevLett.128.175301},
  url = {https://link.aps.org/doi/10.1103/PhysRevLett.128.175301}
}

@article{DeSalvo2015Ultralong,
  title = {Ultra-long-range Rydberg molecules in a divalent atomic system},
  author = {DeSalvo, B. J. and Aman, J. A. and Dunning, F. B. and Killian, T. C. and Sadeghpour, H. R. and Yoshida, S. and Burgd\"orfer, J.},
  journal = {Phys. Rev. A},
  volume = {92},
  issue = {3},
  pages = {031403},
  numpages = {5},
  year = {2015},
  month = {Sep},
  publisher = {American Physical Society},
  doi = {10.1103/PhysRevA.92.031403},
  url = {https://link.aps.org/doi/10.1103/PhysRevA.92.031403}
}

@article{Kolkowitz2015Probing,
author = {S. Kolkowitz  and A. Safira  and A. A. High  and R. C. Devlin  and S. Choi  and Q. P. Unterreithmeier  and D. Patterson  and A. S. Zibrov  and V. E. Manucharyan  and H. Park  and M. D. Lukin },
title = {Probing Johnson noise and ballistic transport in normal metals with a single-spin qubit},
journal = {Science},
volume = {347},
number = {6226},
pages = {1129-1132},
year = {2015},
doi = {10.1126/science.aaa4298},
URL = {https://www.science.org/doi/abs/10.1126/science.aaa4298}}

@article{Casola2018Probing,
	title = {Probing condensed matter physics with magnetometry based on nitrogen-vacancy centres in diamond},
	volume = {3},
	issn = {2058-8437},
	url = {https://doi.org/10.1038/natrevmats.2017.88},
	doi = {10.1038/natrevmats.2017.88},
	number = {1},
	journal = {Nat. Rev. Mater.},
	author = {Casola, Francesco and van der Sar, Toeno and Yacoby, Amir},
	month = jan,
	year = {2018},
	pages = {17088},
}

@article{Ospelkaus2006Interaction,
  title = {Interaction-Driven Dynamics of $^{40}\mathrm{K}\mathrm{\text{\ensuremath{-}}}^{87}\mathrm{Rb}$ Fermion-Boson Gas Mixtures in the Large-Particle-Number Limit},
  author = {Ospelkaus, C. and Ospelkaus, S. and Sengstock, K. and Bongs, K.},
  journal = {Phys. Rev. Lett.},
  volume = {96},
  issue = {2},
  pages = {020401},
  numpages = {4},
  year = {2006},
  month = {Jan},
  publisher = {American Physical Society},
  doi = {10.1103/PhysRevLett.96.020401},
  url = {https://link.aps.org/doi/10.1103/PhysRevLett.96.020401}
}

@article{Chin2010Feshbach,
  title = {Feshbach resonances in ultracold gases},
  author = {Chin, Cheng and Grimm, Rudolf and Julienne, Paul and Tiesinga, Eite},
  journal = {Rev. Mod. Phys.},
  volume = {82},
  issue = {2},
  pages = {1225--1286},
  numpages = {0},
  year = {2010},
  month = {Apr},
  publisher = {American Physical Society},
  doi = {10.1103/RevModPhys.82.1225},
  url = {https://link.aps.org/doi/10.1103/RevModPhys.82.1225}
}

@article{Chen2025Massgap,
  title = {Mass-Gap Description of Heavy Impurities in Fermi Gases},
  author = {Chen, Xin and Dizer, Eugen and Rodr\'{\i}guez, Emilio Ramos and Schmidt, Richard},
  journal = {Phys. Rev. Lett.},
  volume = {135},
  issue = {19},
  pages = {193401},
  numpages = {7},
  year = {2025},
  month = {Nov},
  publisher = {American Physical Society},
  doi = {10.1103/h2f7-dhjh},
  url = {https://link.aps.org/doi/10.1103/h2f7-dhjh}
}

@article{Hoinka2017Goldstone,
  title={Goldstone mode and pair-breaking excitations in atomic {F}ermi superfluids},
  author={Hoinka, Sascha and Dyke, Paul and Lingham, Marcus G and Kinnunen, Jami J and Bruun, Georg M and Vale, Chris J},
  journal={Nature Physics},
  volume={13},
  number={10},
  pages={943--946},
  year={2017},
  publisher={Nature Publishing Group UK London},
  url={https://www.nature.com/articles/nphys4187}
}

@article{PhysRevLett.128.100401,
  title = {Excitation Spectrum and Superfluid Gap of an Ultracold {F}ermi Gas},
  author = {Biss, Hauke and Sobirey, Lennart and Luick, Niclas and Bohlen, Markus and Kinnunen, Jami J. and Bruun, Georg M. and Lompe, Thomas and Moritz, Henning},
  journal = {Phys. Rev. Lett.},
  volume = {128},
  issue = {10},
  pages = {100401},
  numpages = {7},
  year = {2022},
  month = {Mar},
  publisher = {American Physical Society},
  doi = {10.1103/PhysRevLett.128.100401},
  url = {https://link.aps.org/doi/10.1103/PhysRevLett.128.100401}
}

@article{Fey2020Ultralong,
author = {Christian Fey and Frederic Hummel and Peter Schmelcher},
title = {Ultralong-range Rydberg molecules},
journal = {Molecular Physics},
volume = {118},
number = {2},
pages = {e1679401},
year = {2020},
publisher = {Taylor \& Francis},
doi = {10.1080/00268976.2019.1679401},

}

@article{Eiles2019Trilobites,
doi = {10.1088/1361-6455/ab19ca},
url = {https://dx.doi.org/10.1088/1361-6455/ab19ca},
year = {2019},
month = {may},
publisher = {IOP Publishing},
volume = {52},
number = {11},
pages = {113001},
author = {Eiles, Matthew T},
title = {Trilobites, butterflies, and other exotic specimens of long-range Rydberg molecules},
journal = {J. Phys. B}
}

\clearpage

\title{Supplemental Material: Probing Bardeen--Cooper--Schrieffer Pairing and Quasiparticle Formation in Ultracold Gases by Rydberg Atom Spectroscopy}

%
%

\date{\today}

\maketitle

\setcounter{equation}{0}
\setcounter{figure}{0}
\setcounter{table}{0}
\setcounter{page}{1}
\makeatletter
\renewcommand{\theequation}{S\arabic{equation}}
\renewcommand{\thefigure}{S\arabic{figure}}
\renewcommand{\bibnumfmt}[1]{[S#1]}
\renewcommand{\citenumfont}[1]{S#1}

In this supplementary material, we present the procedure to obtain the absorption spectrum using the functional determinant approach (FDA), provide details on the interpretation of the absorption spectra discussed in the main text, and discuss the calculation of quasiparticle weights. For this, in~\Sec{sec:absorption}, we introduce the absorption spectrum and its relation with the Ramsey signal. In~\Sec{sec:Klich}, we introduce the FDA to obtain the Ramsey signal for a one- and two-component Fermi gas. Next, in~\Sec{sec:numerics}, we give an overview of our numerical parameters. In~\Sec{sec:peaks}, we discuss the peaks associated with the weakly bound states and provide a full characterization of the peaks shown in the spectra from the main text. Finally, in~\Sec{sec:qw}, we provide insight on the calculation of the quasiparticle weight.   

\section{Absorption Spectrum}\label{sec:absorption}
Considering the procedure explained in the main text, the impurity is excited from an initial state $|0\rangle$ to the orthogonal Rydberg state $|\mathrm{R}\rangle$. In linear response theory, the absorption spectrum is obtained through Fermi's golden rule,
\begin{equation}
	A(\omega)=2\pi\sum_{f,i}\rho_i|\langle\psi_f|\hat{\Omega}|\psi_i\rangle|^2\delta(\omega-[E_f-E_i]),
\end{equation}
with $\rho_i$ the full many-body density matrix of the initial state and $\hat{\Omega}=|\mathrm{R}\rangle\langle0|+\mathrm{h.c.}$, the transition operator. The sum goes through all the eigen-functions of the initial and final states with the form
\begin{alignat}{4}
	&|\psi_i\rangle&&=|0\rangle\otimes|i\rangle,\quad     &&\hat{H}|\psi_i\rangle&&=E_i|\psi_i\rangle,\\
	&|\psi_f\rangle&&=|\mathrm{R}\rangle\otimes|f\rangle,\quad     &&\hat{H}|\psi_f\rangle&&=E_f|\psi_f\rangle,
\end{alignat}
with $\hat{H}=|\mathrm{0}\rangle\langle\mathrm{0}|\otimes\hat H_{0}+|\mathrm{R}\rangle\langle\mathrm{R}|\otimes\hat{H}_{\mathrm{R}}$ as shown in Eq.~\eqref{eq:hamiltonian} of the main text. 

Thus, the absorption spectrum can be expressed as
\begin{align*}
	A(\omega)=&\ 2\pi\sum_{f,i}\rho_i\langle\psi_i|\hat{\Omega}|\psi_f\rangle\langle\psi_f|\hat{\Omega}|\psi_i\rangle\delta(\omega-[E_f-E_i])\\
	(i)\quad=&\int_{-\infty}^{\infty}\dd t\,\sum_{f,i}\rho_i\langle\psi_i|\ee^{\ii E_it}\hat{\Omega}|\psi_f\rangle\langle\psi_f|\ee^{-\ii E_ft}\hat{\Omega}|\psi_i\rangle  \ee^{\ii \omega t}\\
	(ii)\quad=&\int_{-\infty}^{\infty}\dd t\,\sum_{f,i}\langle i|\hat{\rho}\ \ee^{\ii\hat{H}_0t}|f\rangle\langle f|\ee^{-\ii \hat{H}_{\mathrm{R}}t}|i\rangle \ee^{\ii\omega t}\\
	(iii)\quad=&\int_{-\infty}^{\infty}\dd t\,\mathrm{Tr}\left\{\hat{\rho}\ \ee^{\ii\hat{H}_0t}\ee^{-\ii\hat{H}_\mathrm{R}t}\right\}\ee^{\ii\omega t}\\
	(iv)\quad=&\ 2\,\mathrm{Re}\left\{\int_{0}^{\infty}\dd t\,S(t) \ee^{\ii\omega t}\right\}.
\end{align*}
These steps involve the following:
\begin{enumerate}
	\item[(i)] The Fourier expansion of the delta distribution is implemented to express the spectrum in terms of a time integral.
	\item[(ii)] The operator form of the exponential terms is recovered in terms of the Hamiltonian $\hat{H}$. The degrees of freedom of the impurity are contracted, reducing the summation over the final states to identity.
	\item[(iii)] For the initial states, the summation corresponds to a trace over the bath's degrees of freedom, yielding the Ramsey signal $S(t)$.
	\item[(iv)] Using $S(-t)=S^*(t)$,  we obtain the form given in~\Eq{eq:absorption} of the main text.
\end{enumerate}

\section{Functional determinant approach}\label{sec:Klich}
After establishing the relationship between the absorption spectrum and the Ramsey signal, it is possible to take advantage of the FDA to calculate the trace of the many-body operators. To this end, Klich's formula \cite{klich2003elementary} allows the calculation of the trace $\mathrm{Tr}(\ee^{\hat X})$, where $\hat X$ is a bilinear many-body operator,
\begin{align}\label{eq:bilinear}
	\hat X = \sum_{i,j}\langle i|\hat x|j\rangle\hat c_i^\dagger\hat c_j.
\end{align}
Here, $\hat x$ is the corresponding single-particle operator. When $\hat c_i^\dagger$, $\hat c_j$ are fermionic operators, Klich's formula takes the form:
\begin{align}\label{eq:klich}
	\mathrm{Tr}\left\{\ee^{\hat{X}_1}...\ \ee^{\hat{X}_n}\right\} = \det(\hat\doubleI + \ee^{\hat{x}_1}...\ \ee^{\hat{x}_n}).
\end{align}
We will first give an overview of how to apply Klich's formula to a one-component fermionic gas, and then we will generalize this procedure to a BCS state.

\subsection{One-component Fermi gas}

In this section, we calculate the Ramsey signal for a one-component Fermi gas (see also supplemental material of Ref.~\cite{Gievers2024Probing}). Since we later introduce the BCS state, it is useful to keep track of the two spin components $\sigma = \uparrow, \downarrow$ already here.

The Ramsey signal for the Fermi gas is 
\begin{equation}\label{eq:ramseytr}
	S_{\sigma}(t)=\frac{\mathrm{Tr}\left\{\ee^{-\beta\hat{H}_\sigma}\ee^{\ii\hat{H}_\sigma t}\ee^{-\ii\hat{H}_{\sigma,\mathrm{R}}t}\right\}}{\mathrm{Tr}\left\{\ee^{-\beta\hat{H}_\sigma}\right\}},
\end{equation}
with the Hamiltonians
\begin{equation}
	\hat{H}_{\sigma}=\int_{\bm{r}}\hat{c}^{\dagger}_{\sigma\bm{r}}\hat{h}_{\sigma}\hat{c}_{\sigma\bm{r}},\quad \hat{H}_{\sigma,\mathrm{R}}=\int_{\bm{r}}\hat{c}^{\dagger}_{\sigma\bm{r}}\hat{h}_{\sigma,\mathrm{R}}\hat{c}_{\sigma\bm{r}}.
\end{equation}
Applying~\Eq{eq:klich} to both traces in~\Eq{eq:ramseytr} yields 
\begin{equation}\label{eq:ramseydet}
	S_{\sigma}(t)=\det(\hat\doubleI-\hat{n}(\hat{h}_\sigma) +\hat{n}(\hat{h}_\sigma) \ee^{\ii\hat{h}_\sigma t}\ee^{- \ii\hat{h}_{\sigma,\mathrm{R}}t}).
\end{equation}
Here, $\hat{n}(\hat{h}_\sigma)$ is the Fermi-distribution operator so that we have
\begin{equation}\nonumber
	\hat{n}(\hat{h}_\sigma)=\frac{\hat{\mathds{1}}}{\ee^{\beta\hat{h}_\sigma}+\hat{\mathds{1}}},\quad \mathrm{Tr}\left\{\ee^{-\beta\hat{H}_0}\right\}=\frac{1}{\det(\hat{\mathds{1}}-\hat{n}(\hat{h}_\sigma))}.
\end{equation}
The single-particle Schrödinger equations corresponding to the Hamiltonians $\hat h_\sigma$ and $\hat h_{\sigma,\mathrm{R}}$ are solved by the orthonormal bases $|\bm n_\sigma\rangle$ and $|\bm \alpha_\sigma\rangle$,
\begin{subequations}
	\label{eq:|n>,|alpha>}
	\begin{align}
		\hat{h}_{\sigma}|\bm{n}_{\sigma}\rangle=\xi_{\sigma\bm{n}}|\bm{n}_{\sigma}\rangle,&\quad \sum_{\bm{n}}|\bm{n}_{\sigma}\rangle\langle\bm{n}_{\sigma}|=\hat{\mathds{1}},\label{eq:bfree}\\
		\hat{h}_{\sigma,\mathrm{R}}|\bm{\alpha}_{\sigma}\rangle=\xi_{\sigma\bm{\alpha}}|\bm{\alpha}_{\sigma}\rangle,&\quad \sum_{\bm{\alpha}}|\bm{\alpha}_{\sigma}\rangle\langle\bm{\alpha}_{\sigma}|=\hat{\mathds{1}}.
		\label{eq:brydberg}
	\end{align}
\end{subequations}
The eigenstates are characterized by the quantum numbers $\bm{n}=(n,\ell,m)$ and $\bm{\alpha} =(\alpha,\ell,m)$ where $n$ and $\alpha$ label the radial quantum numbers. Due to the spherical symmetry of the Rydberg potential, the angular-momentum quantum numbers $\ell$ and $m$ are the same for both systems. The eigenenergies are $\xi_{\sigma\bm{n},\bm{\alpha}}=\epsilon_{\sigma\bm{n},\bm{\alpha}}-\mu_{\sigma}$. By inserting the identities from Eq.~\eqref{eq:|n>,|alpha>} (acting on the subspace of spin $\sigma$) in~\Eq{eq:ramseydet}, we obtain
\begin{align}
	\nonumber [\hat\doubleI-\hat{n}(\hat{h}_\sigma)]\hat{\mathds{1}} +\hat{n}(\hat{h}_\sigma) \ee^{\ii\hat{h}_\sigma t}\hat{\mathds{1}}\ee^{-\ii\hat{h}_{\sigma,\mathrm{R}}t}\hat{\mathds{1}}\hat{\mathds{1}}\\
	=\sum_{\bm{n}\bm{n}'}S^{\sigma}_{\bm{nn}'}(t)|\bm{n}_{\sigma}\rangle\langle\bm{n}'_{\sigma}|,
\end{align}
with
\begin{align}
	S^{\sigma}_{\bm{nn}'}(t)=&[1-n(\xi_{\sigma\bm{n}})]\delta_{\bm{n}\bm{n}'}\\
	&+\sum_{\bm{\alpha}}n(\xi_{\sigma\bm{n}})\ee^{\ii\epsilon_{\sigma\bm{n}}t}\langle\bm{n}_{\sigma}|\bm{\alpha}_{\sigma}\rangle\ee^{-\ii\epsilon_{\sigma\bm{\alpha}}t}\langle\bm{\alpha}_{\sigma}|\bm{n}'_{\sigma}\rangle.\nonumber
\end{align}
This results in the already known one-component Ramsey signal:
\begin{equation}
	S_{\sigma}(t)=\det[S^{\sigma}_{\bm{n}\bm{n}'}(t)].
\end{equation}
Consequently, the many-body problem has been reduced to solving the Schr\"odinger equation for the single-particle picture, i.e., the eigenvalue problem of Eqs.~\eqref{eq:bfree}--\eqref{eq:brydberg} and computing the overlaps $\langle\bm{n}_{\sigma}|\bm{\alpha}_{\sigma}\rangle$ of spin $\sigma$.

For a noninteracting two-component Fermi gas, the two spin components $\sigma=\uparrow,\downarrow$ are decoupled and thus the Ramsey signal reduces to the product of the individual signals:
\begin{equation}
	S(t)\stackrel{\Delta=0}{=}S_{\uparrow}(t)S_{\downarrow}(t).
\end{equation}

\subsection{Two-component BCS state}

In order to apply Klich's formula to the BCS state, it is necessary to obtain the diagonalized bilinear form of the many-body Hamiltonians, similar to~\Eq{eq:bilinear}. Note that similar expressions were derived in an FDA analysis of a local impurity immersed in a BCS state~\cite{benjamin2013quasiparticletheoryresonantinelastic,Wang2022Heavy,Wang2022Exact}. We first transform the Hamiltonians in Eqs.~\eqref{eq:hfree} and~\eqref{eq:hrydberg} to the momentum basis that diagonalizes each respective single-particle operator. Using the basis introduced in~\Eq{eq:bfree}, $\hat{H}_0$ takes the form
\begin{subequations}
	\begin{align}
		\hat{H}_0&=\sum_{\bm{n}}\hat{c}^{\dagger}_{\bm{n}}\begin{pmatrix}
			\xi_{\uparrow\bm{n}}&\Delta\\
			\Delta &-\xi_{\downarrow\bm{n}}
		\end{pmatrix}\hat{c}_{\bm{n}}+E_g,\\
		E_g&=\sum_{\bm{n}}\xi_{\downarrow\bm{n}}-\frac{V}{g}|\Delta|^2,
		\label{eq:E_g}
	\end{align}
\end{subequations}
where we use the spinor representation ${\hat{c}^\dagger_{\bm{n}}=\left(\hat{c}^\dagger_{\uparrow\bm{n}},\, \hat{c}_{\downarrow-\bm{n}}\right)}$. Here, $\hat c_{\downarrow,-\bm{n}}$ annihilates a state with flipped momentum sign, i.e., $\bm{k}\to-\bm{k}$, compared to that corresponding to $\hat c_{\downarrow,\bm{n}}$. The two operators are related by the parity identity $\hat c_{\sigma-\bm{n}}=(-1)^\ell\hat c_{\sigma\bm{n}}$. In Eq.~\eqref{eq:E_g}, we show the explicit form of $E_g$, in which the first term arises from the spinor notation. The second term in $E_g$ comes from the mean-field approximation with $g$ the bare interaction between the components of the bath and $V$ a volume factor. Since the Rydberg atom does not equilibrate with the medium, it is an irrelevant term that will not be considered from now on. 

We obtain the form of \Eq{eq:bilinear} with the help of the Bogoliubov transformation, considering the Bogoliubov operators:
\begin{equation}\label{eq:bogol}
	\hat{\gamma}_{\bm{n}}=\begin{pmatrix}
		\hat{\gamma}_{\uparrow\bm{n}}\\[0.2em]
		\hat{\gamma}^\dagger_{\downarrow-\bm{n}}
	\end{pmatrix}\equiv\begin{pmatrix}
		\phantom{-}u_{\bm{n}}&v_{\bm{n}}\\
		-v_{\bm{n}}&\phantom{-}u_{\bm{n}}
	\end{pmatrix}\hat{c}_{\bm{n}},\quad |u_{\bm{n}}|^2+|v_{\bm{n}}|^2=1.
\end{equation}
Here, $u_{\bm{n}}$ and $\ v_{\bm{n}}$ are real-valued functions. As we consider spherically symmetric potentials, we have $u_{\bm{n}}=u_{-\bm{n}}$, $v_{\bm{n}}=v_{-\bm{n}}$ 
It follows from applying the transformation with the normalization condition that
\begin{align}\label{eq:bcscoef}
	\begin{pmatrix}
		u_{\bm{n}}^2\\[0.4em]\ v_{\bm{n}}^2
	\end{pmatrix}=\frac{1}{2}\left(1\pm\frac{\overline{\xi}_{\bm{n}}}{\sqrt{\overline{\xi}_{\bm{n}}^2+\Delta^2}}\right),\quad \overline{\xi}_{\bm{n}}=\frac{\xi_{\uparrow\bm{n}}+\xi_{\downarrow\bm{n}}}{2}.
\end{align}

\begin{figure}
	\centering
	\includegraphics[width=0.4\textwidth]{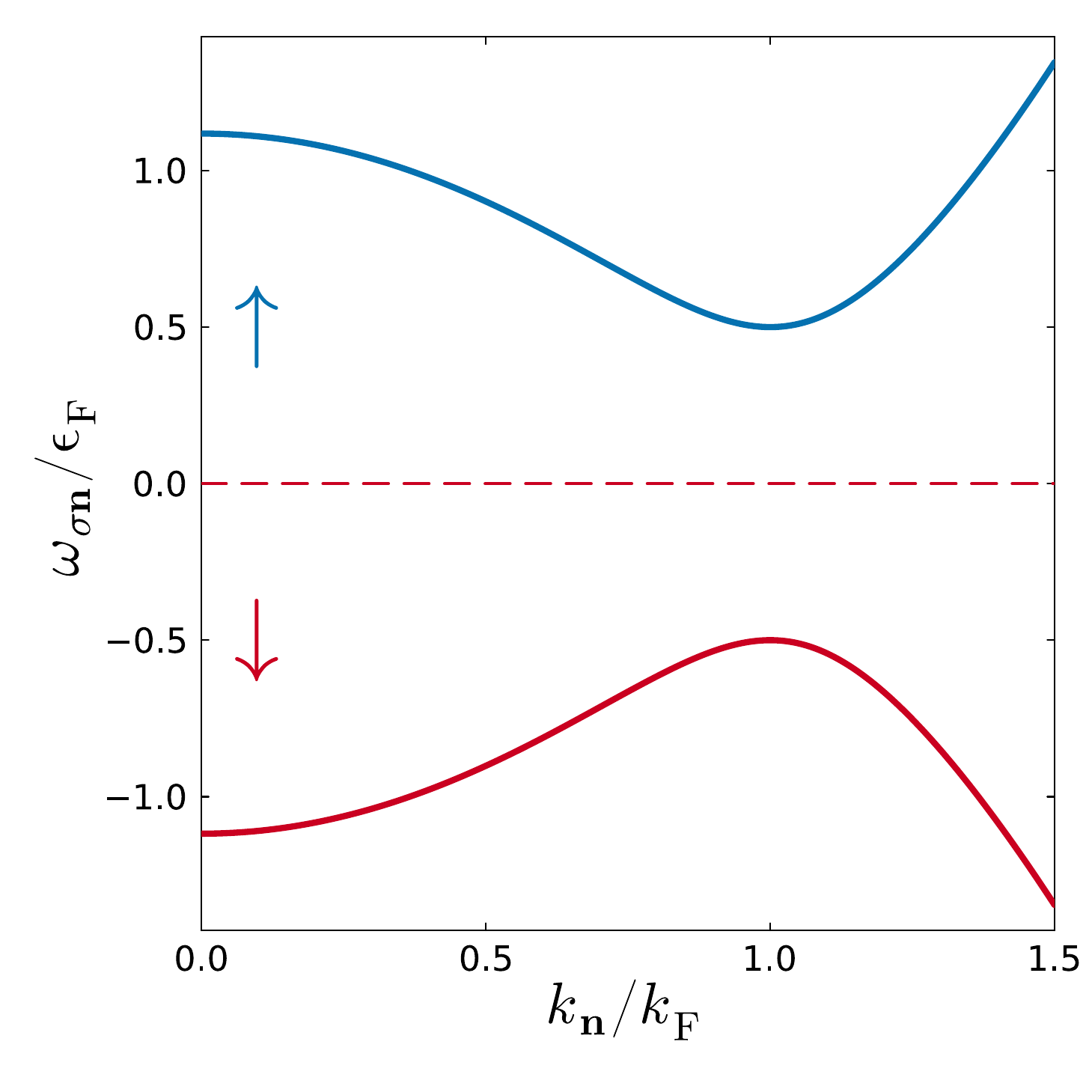}\hfill\\
	\vspace{-.5em}
	\caption{Bogoliubov dispersion for a balanced system, i.e., $m_{\uparrow}=m_{\downarrow}=m$, $\mu_{\uparrow}=\mu_{\downarrow}=\mu$ and $\omega_{\uparrow\bm{n}}=-\omega_{\downarrow\bm{n}}$, for ${\Delta/\epsilon_{\mathrm{F}}=0.5}$. At $T=0$ only the lower branch ($\downarrow$) is occupied, there can be a  Bogoliubov excitation of $2\Delta/\epsilon_\mathrm{F}$ from the lower to the upper branch.}
	\label{fig:bcs_dispersion}
\end{figure}

This results in the diagonal Hamiltonian:
\begin{equation}
	\hat{H}_0=\sum_{\bm{n}}\left[\omega_{\uparrow\bm{n}}\hat{\gamma}^\dagger_{\uparrow\bm{n}}\hat{\gamma}_{\uparrow\bm{n}}+\omega_{\downarrow\bm{n}}\hat{\gamma}_{\downarrow\bm{n}}\hat{\gamma}^\dagger_{\downarrow\bm{n}}+\xi_{\downarrow\bm{n}}\right],
\end{equation}
with the gapped dispersion relation (cf.\ \Fig{fig:bcs_dispersion})
\begin{equation}\label{eq:bcsdispersion}
	\omega_{\sigma\bm{n}}=\frac{\xi_{\uparrow\bm{n}}-\xi_{\downarrow\bm{n}}}{2}+\sigma\sqrt{\overline{\xi}_{\bm{n}}^2+\Delta^2}.
\end{equation}
Note that $\sigma$ on the right-hand side denotes the sign $+,-$ corresponding to the spin index $\sigma=\uparrow,\downarrow$ on the left-hand side.
The corresponding single-particle operator and eigen-basis are written as
\begin{equation}
	\hat{h}_{\Delta}|\Delta_{\sigma\bm{n}}\rangle=\omega_{\sigma\bm{n}}|\Delta_{\sigma\bm{n}}\rangle,\quad \sum_{\sigma\bm{n}}|\Delta_{\sigma\bm{n}}\rangle\langle\Delta_{\sigma\bm{n}}|=\hat{\mathds{1}}\label{eq:bbcs}.
\end{equation}
Here, the single-particle eigenstates $|\Delta_{\sigma\bm{n}}\rangle$ of $\hat h_\Delta$ are linear combinations of the original eigenstates ${|\pm\bm{n}_\sigma\rangle}$:
\begin{align}
	\left(\begin{matrix}|\Delta_{\uparrow\bm{n}}\rangle\\ |\Delta_{\downarrow,-\bm{n}}\rangle\end{matrix}\right)
	&= \left(\begin{matrix}
		\phantom{-}u_{\bm{n}} & v_{\bm{n}} \\
		-v_{\bm{n}} & u_{\bm{n}} \end{matrix}\right)\left(\begin{matrix} |\bm{n}_\uparrow\rangle  \\ |-\bm{n}_\downarrow\rangle\end{matrix}\right).
	\label{eq:Bogoliubov_states_1}
\end{align}
This equation can be compactified as
\begin{align}    |\Delta_{\sigma\bm{n}}\rangle=u_{\bm{n}}|\bm{n}_{\sigma}\rangle + \sigma v_{\bm{n}}|-\bm{n}_{-\sigma}\rangle,   \label{eq:Bogoliubov_states}
\end{align}
where $\sigma$ is used as a symbol for $\uparrow,\downarrow$ and $+,-$ with a flipped spin $\sigma$ denoted as $-\sigma$. 
Note that due to the properties of $|\bm{n}_\sigma\rangle$, Eq.~\eqref{eq:bfree}, and the Bogoliubov transformation, Eq.~\eqref{eq:bogol}, the new eigenstates are orthonormal, i.e., $\langle\Delta_{\sigma\bm{n}}|\Delta_{\sigma'\bm{n}'}\rangle=\delta_{\sigma\sigma'}\delta_{\bm{n}\bm{n}'}$.

Applying the analogous procedure to $\hat{H}$ with the basis in \Eq{eq:brydberg}, i.e., changing $\bm{n}\rightarrow\bm{\alpha}$, results in the diagonalized form
\begin{equation}
\hat{H}=\sum_{\bm{\alpha}}\left[\omega_{\uparrow\bm{\alpha}}\hat{\gamma}^\dagger_{\uparrow\bm{\alpha}}\hat{\gamma}_{\uparrow\bm{\alpha}}+\omega_{\downarrow\bm{\alpha}}\hat{\gamma}_{\downarrow\bm{\alpha}}\hat{\gamma}^\dagger_{\downarrow\bm{\alpha}}+\xi_{\downarrow\bm{\alpha}}\right],
\end{equation}
and the single-particle representation
\begin{subequations}
\label{eq:bbcsrydberg}
\begin{align}
	\hat{h}_{\Delta,\mathrm{R}}|\Delta_{\sigma\bm{\alpha}}\rangle&=
	\omega_{\sigma\bm{\alpha}}|\Delta_{\sigma\bm{\alpha}}\rangle,\quad \sum_{\sigma\bm{\alpha}}|\Delta_{\sigma\bm{\alpha}}\rangle\langle\Delta_{\sigma\bm{\alpha}}|=\hat{\mathds{1}},\\
	\text{with}\quad |\Delta_{\sigma\bm{\alpha}}\rangle&=u_{\bm{\alpha}}|\bm{\alpha}_{\sigma}\rangle + \sigma v_{\bm{\alpha}}|-\bm{\alpha}_{-\sigma}\rangle.
\end{align}
\end{subequations}
This enables the use of Klich's formula to obtain the Ramsey signal. Following the steps from Sec.~\ref{sec:Klich} results in the expression for the BCS Ramsey signal,
\begin{align}
S(t)=\frac{\mathrm{Tr}\left\{\ee^{-\beta\hat{H}_0}\ee^{\ii\hat{H}_0 t}\ee^{-\ii\hat{H}t}\right\}}{\mathrm{Tr}\left\{\ee^{-\beta\hat{H}_0}\right\}}=\ee^{\ii\delta\xi_0t}\det[S^{\sigma\sigma'}_{\bm{n}\bm{n}'}(t)],
\label{eq:bcsramsey}
\end{align}
with $\delta\xi_0=\sum_{\bm{n}}\xi_{\downarrow\bm{n}}-\sum_{\bm{\alpha}}\xi_{\downarrow\bm{\alpha}}$
and 
\begin{align}\label{eq:BCSmatrix}
&S^{\sigma\sigma'}_{\bm{n}\bm{n}'}(t)=[1-n(\omega_{\sigma\bm{n}})]\delta_{\bm{n}\bm{n}'}\delta_{\sigma\sigma'}\\
&+n(\omega_{\sigma\bm{n}})\ee^{\ii\omega_{\sigma\bm{n}}t}\sum_{\sigma''\bm{\alpha}}\langle\Delta_{\sigma\bm{n}}|\Delta_{\sigma''\bm{\alpha}}\rangle\ee^{-\ii\omega_{\sigma''\bm{\alpha}}t}\langle\Delta_{\sigma''\bm{\alpha}}|\Delta_{\sigma'\bm{n}'}\rangle.\nonumber    
\end{align}
Moreover, we can relate the overlaps $\langle\Delta_{\sigma\bm{n}}|\Delta_{\sigma''\bm{\alpha}}\rangle$ to the one-component overlaps $\langle\bm{n}_{\sigma}|\bm{\alpha}_{\sigma''}\rangle$ using the Bogoliubov transformations shown in~\Eqs{eq:Bogoliubov_states} and \eqref{eq:bbcsrydberg}:
\begin{align}
	\nonumber
	\langle\Delta_{\uparrow\bm{n}}|\Delta_{\uparrow\bm{\alpha}}\rangle &= u_{\bm{n}}u_{\bm{\alpha}}\langle\bm{n}_\uparrow|\bm{\alpha}_\uparrow\rangle + v_{\bm{n}}v_{\bm{\alpha}}\langle-\bm{n}_\downarrow|-\bm{\alpha}_\downarrow\rangle,\\
	\nonumber
	\langle\Delta_{\downarrow\bm{n}}|\Delta_{\downarrow\bm{\alpha}}\rangle &= u_{\bm{n}}u_{\bm{\alpha}}\langle\bm{n}_\downarrow|\bm{\alpha}_\downarrow\rangle + v_{\bm{n}}v_{\bm{\alpha}}\langle-\bm{n}_\uparrow|-\bm{\alpha}_\uparrow\rangle,\\
	\nonumber
	\langle\Delta_{\uparrow\bm{n}}|\Delta_{\downarrow\bm{\alpha}}\rangle &= v_{\bm{n}}u_{\bm{\alpha}}\langle-\bm{n}_\downarrow|\bm{\alpha}_\downarrow\rangle - u_{\bm{n}}v_{\bm{\alpha}}\langle\bm{n}_\uparrow|-\bm{\alpha}_\uparrow\rangle,\\
	\langle\Delta_{\downarrow\bm{n}}|\Delta_{\uparrow\bm{\alpha}}\rangle &= -v_{\bm{n}}u_{\bm{\alpha}}\langle-\bm{n}_\uparrow|\bm{\alpha}_\uparrow\rangle + u_{\bm{n}}v_{\bm{\alpha}}\langle\bm{n}_\downarrow|-\bm{\alpha}_\downarrow\rangle.
	\label{eq:<Delta|Delta>_details}
\end{align}
The one-component overlaps $\langle\bm{n}_{\sigma}|\bm{\alpha}_{\sigma''}\rangle$ are diagonal in $\sigma$, $\ell$ and $m$. The minus signs in $\langle-\bm{n}_\sigma|-\bm{\alpha}_\sigma\rangle$ cancel since $(-1)^{2\ell}=1$, while $\langle-n_\sigma|\alpha_\sigma\rangle$ and $\langle n_\sigma|-\alpha_\sigma\rangle$ produce an additional factor $(-1)^\ell$. Let us compactify Eq.~\eqref{eq:<Delta|Delta>_details} as
\begin{align}
\langle\Delta_{\sigma\bm{n}}|\Delta_{\sigma''\bm{\alpha}}\rangle
&=M^{\sigma}_{\bm{n}\bm{\alpha}}\delta_{\sigma\sigma''}-\sigma (-1)^\ell L^{\sigma}_{\bm{n}\bm{\alpha}}\delta_{\sigma-\sigma''},
\end{align}
with
\begin{subequations}
\label{eq:M_L_matrices}
\begin{align}
	M^{\sigma}_{\bm{n}\bm{\alpha}}&=
	u_{\bm{n}}u_{\bm{\alpha}}
	\langle\bm{n}_{\sigma}|\bm{\alpha}_{\sigma}\rangle+v_{\bm{n}}v_{\bm{\alpha}}\langle\bm{n}_{-\sigma}|\bm{\alpha}_{-\sigma}\rangle,\\ L^{\sigma}_{\bm{n}\bm{\alpha}}&=u_{\bm{n}}v_{\bm{\alpha}}\langle\bm{n}_{\sigma}|\bm{\alpha}_{\sigma}\rangle-v_{\bm{n}}u_{\bm{\alpha}}\langle\bm{n}_{-\sigma}|\bm{\alpha}_{-\sigma}\rangle\, .
\end{align}
\end{subequations}
Inserting this expansion in~\Eq{eq:BCSmatrix} yields the expression
\begin{align}
S^{\sigma\sigma'}_{\bm{n}\bm{n}'}(t)=[1-n(\omega_{\sigma\bm{n}})]&\delta_{\bm{n}\bm{n}'}\delta_{\sigma\sigma'}\nonumber\\
+n(\omega_{\sigma\bm{n}})\ee^{\ii t\omega_{\sigma\bm{n}}}\sum_{\bm{\alpha}}&\left[M^{\sigma}_{\bm{n\alpha}}\ee^{-\ii t\omega_{\sigma\bm\alpha}}M^{\sigma}_{\bm{\alpha}\bm{n}'}\right.\nonumber\\
&\left.-L^{\sigma}_{\bm{n\alpha}}\ee^{-\ii t\omega_{-\sigma\bm\alpha}}L^{-\sigma}_{\bm{\alpha}\bm{n}'}\right]\delta_{\sigma\sigma'}\nonumber\\
-\sigma(-1)^\ell n(\omega_{\sigma\bm{n}})\ee^{\ii t\omega_{\sigma\bm{n}}}\sum_{\bm{\alpha}}&\left[M^{\sigma}_{\bm{n\alpha}}\ee^{-\ii t\omega_{\sigma\bm\alpha}}L^{\sigma}_{\bm{\alpha}\bm{n}'}\right.\nonumber\\
&\left.+L^{\sigma}_{\bm{n\alpha}}\ee^{-\ii t\omega_{-\sigma\bm\alpha}}M^{-\sigma}_{\bm{\alpha}\bm{n}'}\right]\delta_{\sigma-\sigma'}\ .\label{eq:expmatrix}   
\end{align}

For a balanced system, i.e., $\xi_{\uparrow\bm{n},\bm{\alpha}}=\xi_{\downarrow\bm{n},\bm{\alpha}}$, $\langle\bm{n}_\sigma|\bm{\alpha}_\sigma\rangle$ and $\langle\bm{n}_{-\sigma}|\bm{\alpha}_{-\sigma}\rangle$ are identical, making $M_{\bm{n\alpha}}$ and $L_{\bm{n\alpha}}$ independent of $\sigma$. Furthermore, this yields $M_{\bm{\alpha n}}=(M_{\bm{n \alpha}})^*$ and $L_{\bm{\alpha n}}=-(L_{\bm{n \alpha}})^*$ and $\omega_{-\sigma\bm{n}}=-\omega_{\sigma\bm{n}}$. The last identity implies $1-n(\omega_{-\sigma\bm{n}})=n(\omega_{\sigma\bm{n}})$.

In order to compute the full determinant, we can perform a partial determinant of the spin index
\begin{align}
\nonumber\det S^{\sigma\sigma'}&=\det\left(\begin{matrix}S^{\uparrow\uparrow} & S^{\uparrow\downarrow}\\
	S^{\downarrow\uparrow} & S^{\downarrow\downarrow}\end{matrix}\right) \\[0.3em]
&= \det \left[S^{\uparrow\uparrow}S^{\downarrow\downarrow}-S^{\uparrow\uparrow}S^{\downarrow\uparrow}(S^{\uparrow\uparrow})^{-1}S^{\uparrow\downarrow}\right]. \label{eq:det_st}
\end{align}
Note that the factors $(-1)^\ell$ from Eq.~\eqref{eq:expmatrix} exactly cancel in the determinant as $S^{\downarrow\uparrow}$ and $S^{\uparrow\downarrow}$ are multiplied.

Consequently, the two-component interacting many-body problem reduces to computing the single-particle eigenvalues $\xi_{\sigma \bm{n}}$ and $\xi_{\sigma\bm{\alpha}}$, and obtaining the overlaps $\langle\bm{n}_{\sigma}|\bm{\alpha}_{\sigma}\rangle$. This problem is a straightforward numerical calculation of the Schr\"odinger equation. The single-particle wave functions $|\bm{n}_\sigma\rangle$, $|\bm{\alpha}_\sigma\rangle$ as well as their overlaps $\langle\bm{n}_\sigma|\bm{\alpha}_\sigma\rangle$ are computed via the procedure extensively described in the supplemental material of Ref.~\cite{Gievers2024Probing}.

With the one-component eigenvalues, the Bogoliubov energy modes $\left\{\omega_{\sigma\bm{n}},\ \omega_{\sigma\bm{\alpha}}\right\}$ can be obtained through \Eq{eq:bcsdispersion} for the free and interacting cases. Furthermore, with \Eq{eq:bcscoef} and the one-component overlaps, the matrices $M^\sigma_{\bm{n\alpha}}$ and $L^\sigma_{\bm{n\alpha}}$ are determined. Consequently, the Ramsey signal is computed by taking the determinant of \Eq{eq:expmatrix} with the numerical simplification of~\Eq{eq:det_st}.

Observe that the computation of the Ramsey components, $S^{\sigma\sigma'}$, in~\Eq{eq:expmatrix} is evaluated with respect to the occupation of the Bogoliubov modes. At $T=0$, one can further simplify the computation by noticing that only the negative branch is occupied (see \Fig{fig:bcs_dispersion}) such that $n(\omega_{\uparrow\bm{n}})=0$ and $n(\omega_{\downarrow\bm{n}})=1$. Consequently, $\det(S^{\sigma\sigma'})\stackrel{T=0}{=}\det(S^{\downarrow\downarrow})$.

This procedure is done for increasing angular momentum $\ell$ until the contribution to the Ramsey signal converges to unity; the finite-range nature of the potential sets an upper bound for the relevant angular-momentum scattering channels that need to be considered, i.e., ${\ell\lesssim\ell_{\mathrm{max}}}$. In the data presented, we used $\ell_\mathrm{max}=5$.

After obtaining the Ramsey signal for each angular component, $S^\ell(t)$, the total signal is obtained by 
\begin{equation}\label{eq:l_ramsey}
S(t)=\prod_{\ell=0}^{\ell_{\max}}\left[S^{\ell}(t)\right]^{2\ell+1}\ .
\end{equation}

\section{Numerical Calculation}\label{sec:numerics}

\begin{figure}
\centering
\includegraphics[width=0.48\textwidth]{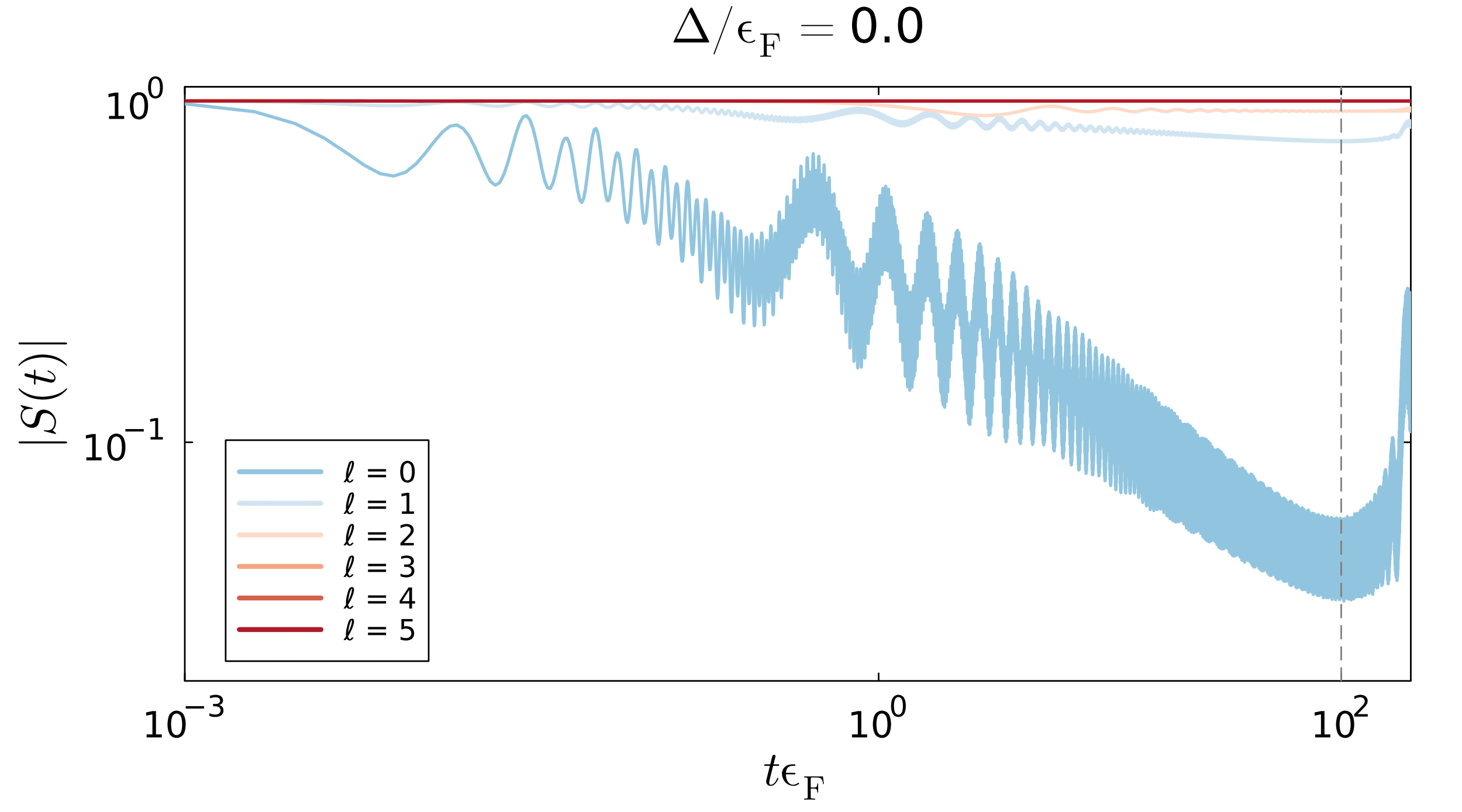}\hfill\\
(a)\hfill\\
\vspace{.5em}
\includegraphics[width=0.48\textwidth]{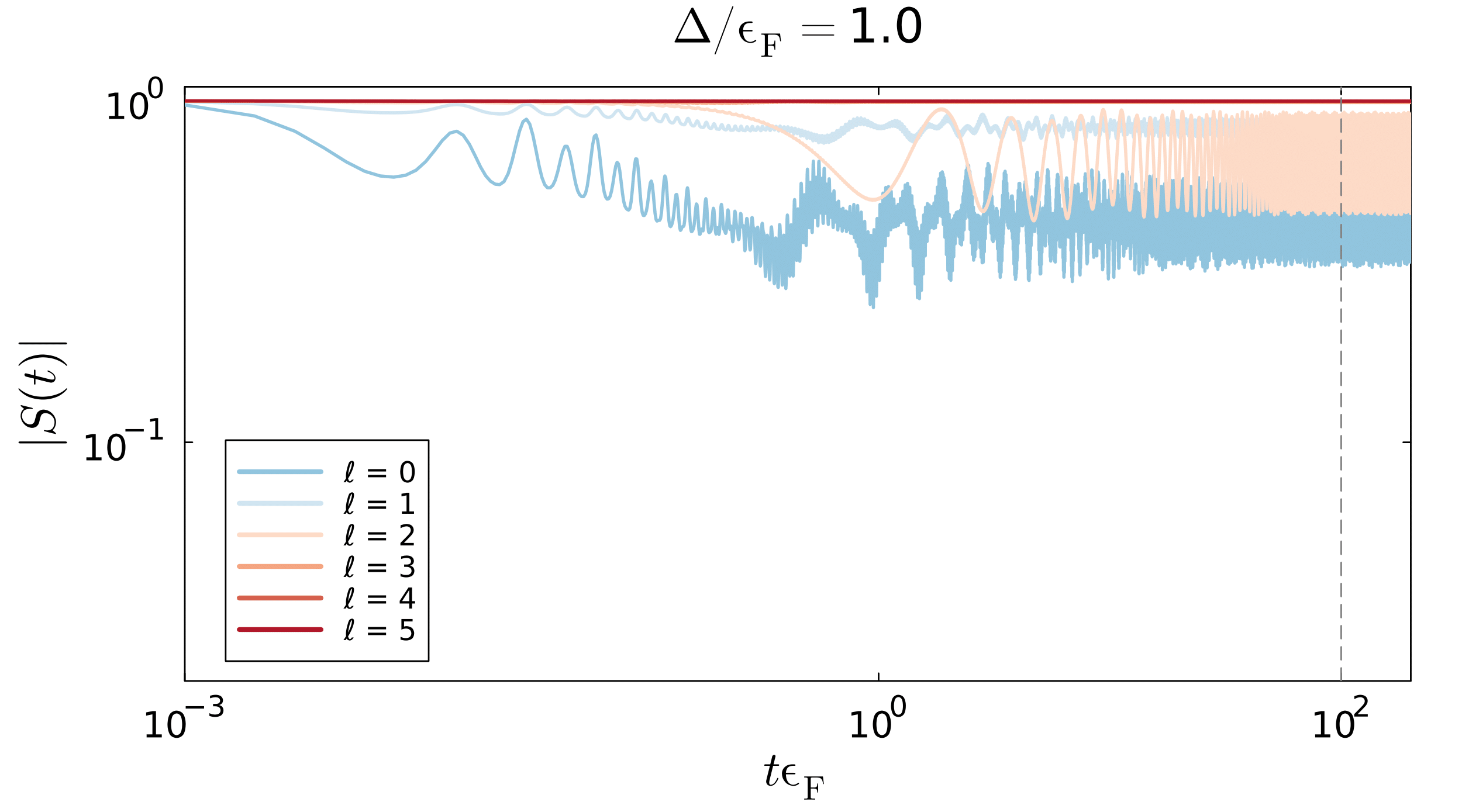}\hfill\\
(b)\hfill\\
\vspace{-.5em}
\caption{Ramsey signal decomposed by contributions for different angular-momentum channels; $\ell_{\max}\approx5$. As the gap increases, higher angular-momentum channels will contribute to the oscillations in the total signal. The dashed line represents the onset of system size effects at time $t_{\max}\approx R/2$.}
\label{fig:l_ramsey}
\end{figure}

In this section, we briefly go through the numerical parameters needed to obtain the absorption spectrum shown in the main text. We use natural units, $\epsilon_{\mathrm{F}}=1=2m_{\sigma}$ and $m_\uparrow=m_\downarrow$. In order to resolve the trimer peak, the time resolution $\delta t$, of the Ramsey signal must satisfy ${E_{T}<2\pi/\delta t=\omega_{\mathrm{max}}}$.

Furthermore, as seen in~\Fig{fig:l_ramsey}(a) for the $s$-wave component, the duration of the Ramsey signal is limited by the size of the system. Here, the signal stops decaying to a vanishing value; as it should do for an infinite mass impurity in a Fermi sea, where $|S(t\rightarrow\infty)|\sim N^{-\alpha}$ with $N$ the number of particles in the Fermi sea~\cite{anderson_infrared_1967,Chen2025Massgap}. These effects become noticeable at times ${t_{\mathrm{max}}\sim R/v_{\mathrm{F}} = R/2}$. In our calculations, we use $k_{\mathrm{F}}R=200$ such that Ramsey signals have a duration of $t\epsilon_\mathrm{F}=100$, and we choose a numerical resolution of $\delta t\,\epsilon_\mathrm{F}=0.001$. All signals are normalized such that $S_{\text{norm}}(t)=S(t)/S(0)$.

Obtaining the Ramsey signal for each angular component (cf.\ \Fig{fig:l_ramsey}), we find convergence to unity for $\ell_{\max}=5$. Note that, as the gap increases, higher angular-momentum channels contribute. This is reflected in the absorption spectrum by the emergence of higher angular-momentum peaks.

\section{Peak Structure: Polaron Resonance}\label{sec:peaks}
\begin{figure}
\centering
\includegraphics[width=0.48\textwidth]{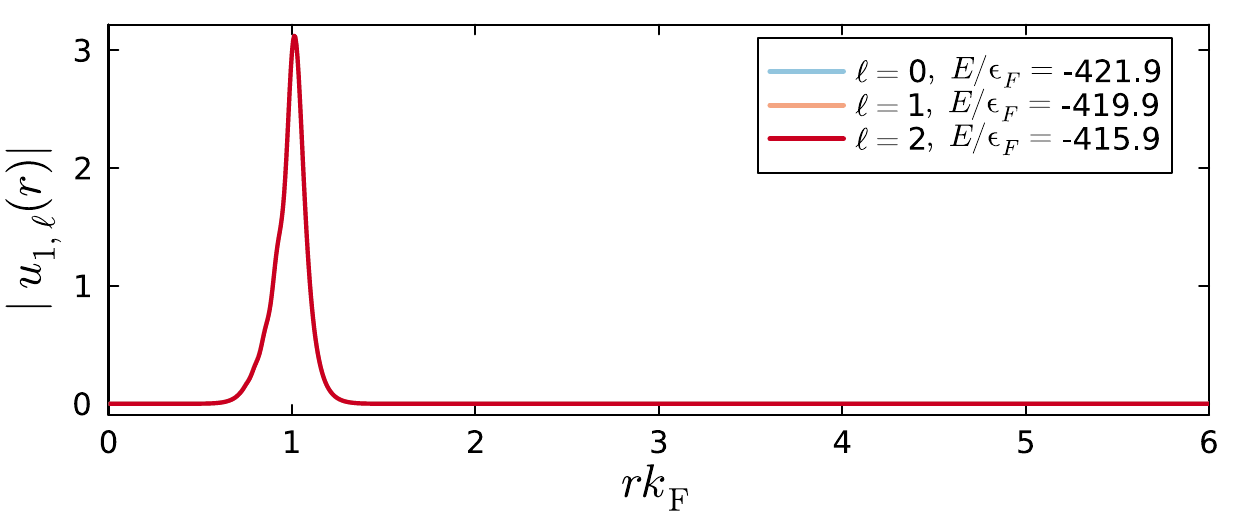}\hfill\\
\includegraphics[width=0.48\textwidth]{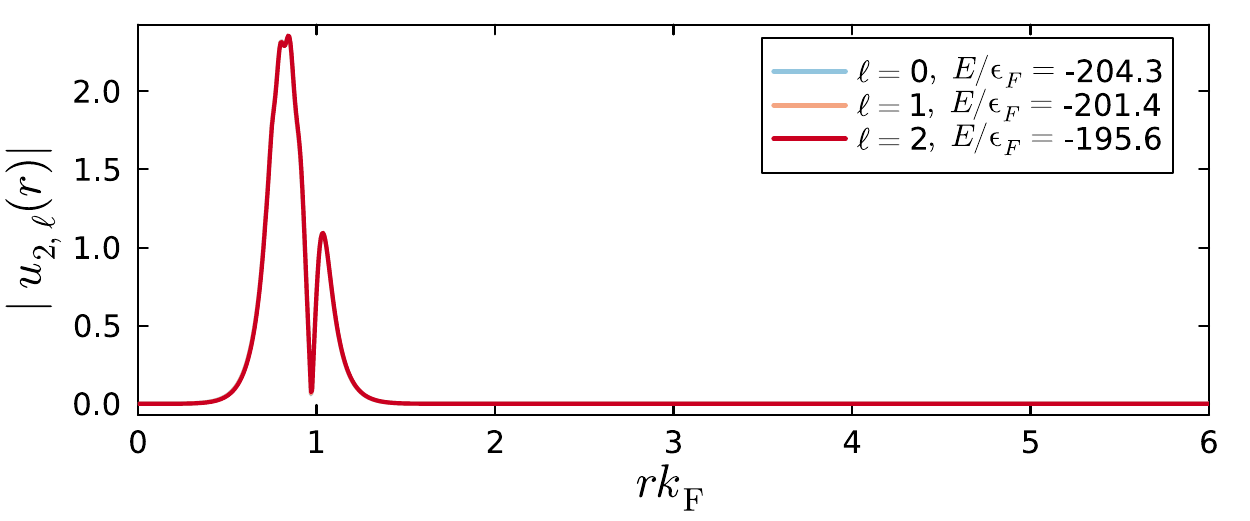}\hfill\\
\includegraphics[width=0.48\textwidth]{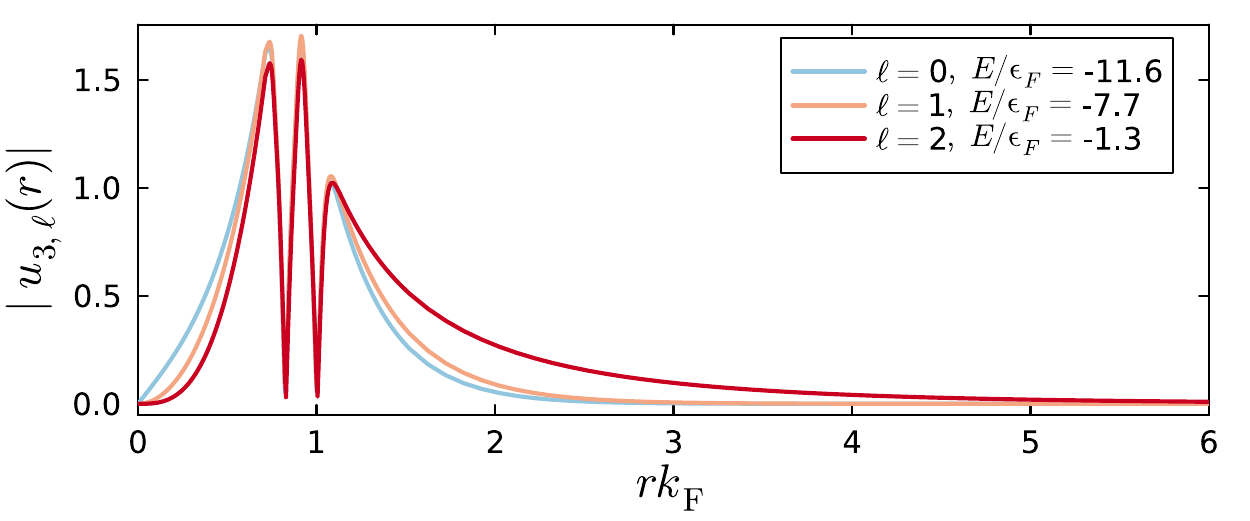}\hfill\\
\caption{Wave functions (radial profile) and binding energies from the deepest (top) to the weakest (bottom) bound state for increasing angular momenta. The relevant bound state $u_{1,\ell}(r)$ with the deepest binding energy (top) is located around the Rydberg radius $r_\mathrm{Ryd}$. The weaker bound states (middle, bottom) are more spread over the space and exhibit a dip at the Rydberg radius.
}
\label{fig:wavefuncs}
\end{figure}
As outlined in the main text, each peak in the spectrum corresponds to a specific physical configuration. Namely, it corresponds to the occupation of bound states originating from the Rydberg potential. The oscillatory Rydberg potential hosts several bound states with radial wave functions denoted by $u_{n_b,\ell}(r)$ (see~\Fig{fig:wavefuncs}). The relevant physics is governed by the deepest bound state, $u_{1,\ell}(r)$, located around the Rydberg radius $r_\mathrm{Ryd}$ (see~\Fig{fig:wavefuncs} top). This is the only bound state considered in the discussion in the main text. In addition, there are several highly nonuniversal weakly bound states that carry spectral weight and contribute to additional peaks in the spectrum (cf.\ Fig.~\ref{fig:spectrum}(a) in the main text). In the following, we discuss how these lead to additional peaks in the vicinity of the dimer and trimer peaks corresponding to the deepest bound state.

\subsection{Weakly bound state}
\begin{figure}
\centering
\includegraphics[width=0.48\textwidth]{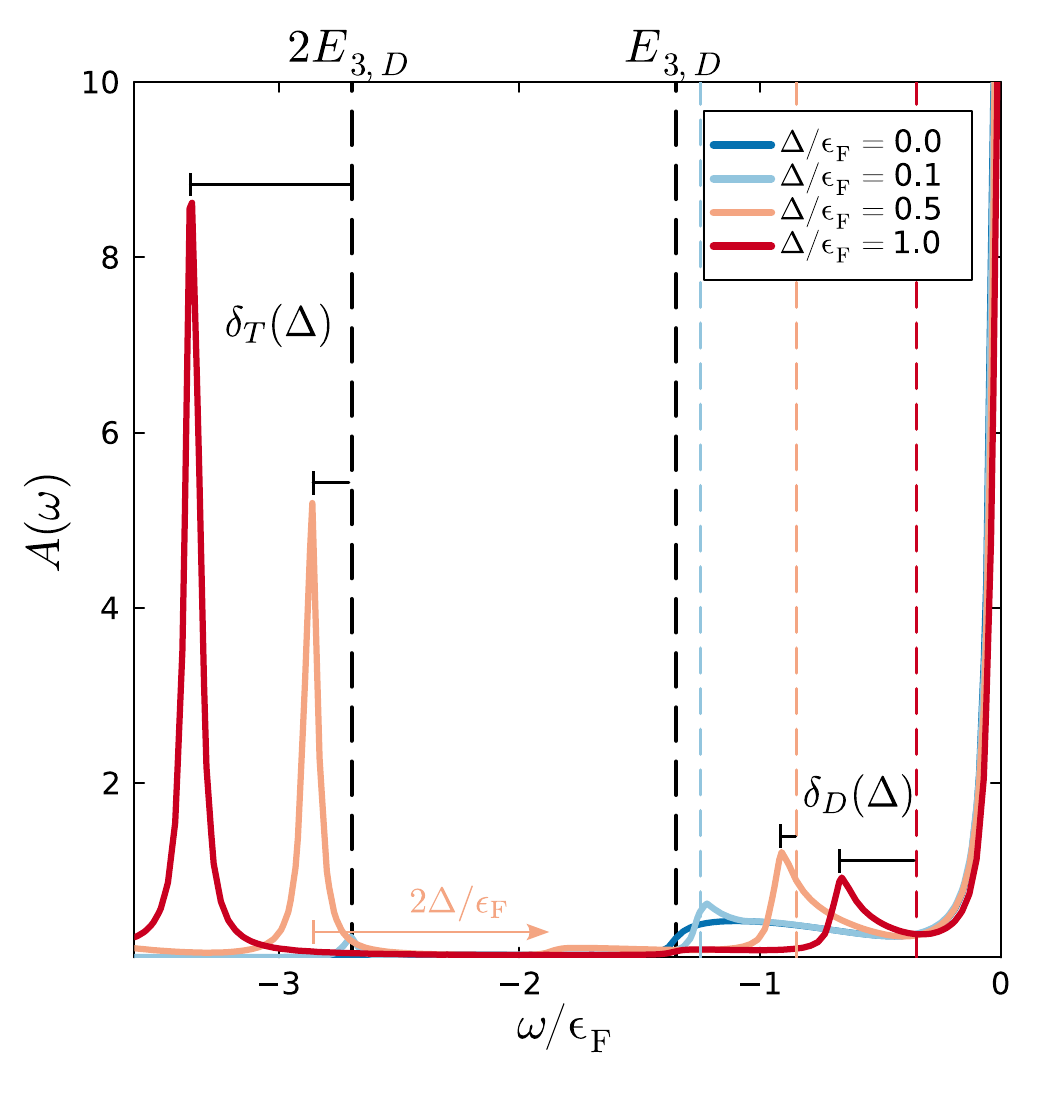}\hfill\\
\vspace{-.5em}
\caption{Full peak structure in the vicinity of the weakest bound state, corresponding to the $\ell=2$ channel with energy $E_{3,D}$, with the dimer and trimer ($2E_{3,D}$) configurations (dashed black lines) and the colored dashed lines denoting the relation $E_{3,D}+\Delta$. Also shown are the anomalous shifts $\delta_D(\Delta)$, Eq.~\eqref{eq:anom_dimer}, and $\delta_T(\Delta)$, Eq.~\eqref{eq:anom_trimer}, for increasing gap strengths. For completeness, the Bogoliubov excitation of the trimer peak for $\Delta/\epsilon_F=0.5$ is visible and marked by an arrow.
}
\label{fig:anomalous}
\end{figure}
As outlined in the main text, the dimer peaks originate from the single occupation of a bound state and, for increasing gap strengths, the peaks shift by ${\Delta/\epsilon_\mathrm{F}}$. Figure~\ref{fig:anomalous} shows the peak structure in the vicinity of the weakest bound state, $n_b=3$, with an angular momentum of ${\ell=2}$, i.e., the bound state with the highest in energy with ${E_{3,D}=-1.3\,\epsilon_\mathrm{F}}$ (cf.\ \Fig{fig:wavefuncs} bottom). Here, the shift in the dimer peak deviates from \Eq{eq:gap} for increasing gap strengths. Consequently, the energy of this weak dimer state for higher gap strengths, denoted by $E^{(\Delta)}_{3,D}$, follows  
\begin{equation}\label{eq:anom_dimer}
E^{(\Delta)}_{3,D}\approx E_{3,D}+\Delta-\delta_D(\Delta),
\end{equation}
where $\delta_{D}(\Delta)$ is a nonlinear correction due to the interplay between the gap parameter $\Delta$ and the angular momentum $\ell$, which we will refer to as the \emph{anomalous} shift. In our case, this effect becomes noticeable for $\Delta/\epsilon_\mathrm{F}>0.2$ and $\ell= 2$.

Furthermore, we also notice a nonlinear correction for the corresponding weak trimer state ${E_{3,T}=2E_{3,D}}$:
\begin{equation}\label{eq:anom_trimer}
E^{(\Delta)}_{3,T}\approx E_{3,T}-\delta_{T}(\Delta)\ ,
\end{equation}
which takes a different value than $\delta_D(\Delta)$ and is not discernible around trimer peaks corresponding to deeper bound states and lower angular momenta.

An exact prediction of the anomalous spectral shifts would require the study of the molecular wave function associated with this weakly bound state, which lies beyond the scope of this work. Nevertheless, it is possible to obtain an intuition why these peaks deviate from the dimer peak behavior of the other deeply bound states.

To this end, we analyze the bound states wave functions for increasing angular momenta, as shown in~\Fig{fig:wavefuncs}. For the deeply bound states, the radial profiles remain essentially unchanged for higher partial waves. In contrast, for the weak bound state corresponding to $E_{3,D}$ (\Fig{fig:wavefuncs} bottom), the probability distribution varies when increasing the angular momentum. Due to the resonant energy scale, $E_{3,D}\sim\epsilon_\mathrm{F}$ where there is no clear separation of physical phenomena, the weakly bound state has enhanced scattering with the continuum, reminiscent of the strongly interacting polaron problem where the lifetime of the resonant level leads to a smooth profile in the spectrum~\cite{Combescot1971Infrared}. Increasing the gap strength regulates this resonant scattering with the continuum, thus recovering the structure and decaying characteristics of the dimer peak for increasing gap strengths.  

Characterizing these shifts for the weakly bound state, we can obtain all configurations for the peaks seen in the dimer and trimer profiles corresponding to the deepest bound state [cf.\ \Fig{fig:spectrum}(b)--(c)]. Furthermore, the potential can be tuned so that the weakest bound state does not appear~\cite{greene2000creation}. Since we can characterize the peaks arising from it, our analysis for the deepest bound states remains a general result. In the following sections, we analyze the peak structure arising from these weakly bound state specifically for the $\Delta=1\epsilon_\mathrm{F}$ spectrum, where the anomalous shift has the most noticeable impact.

\subsection{Dimer}
\begin{figure}
\centering
\includegraphics[width=0.48\textwidth]{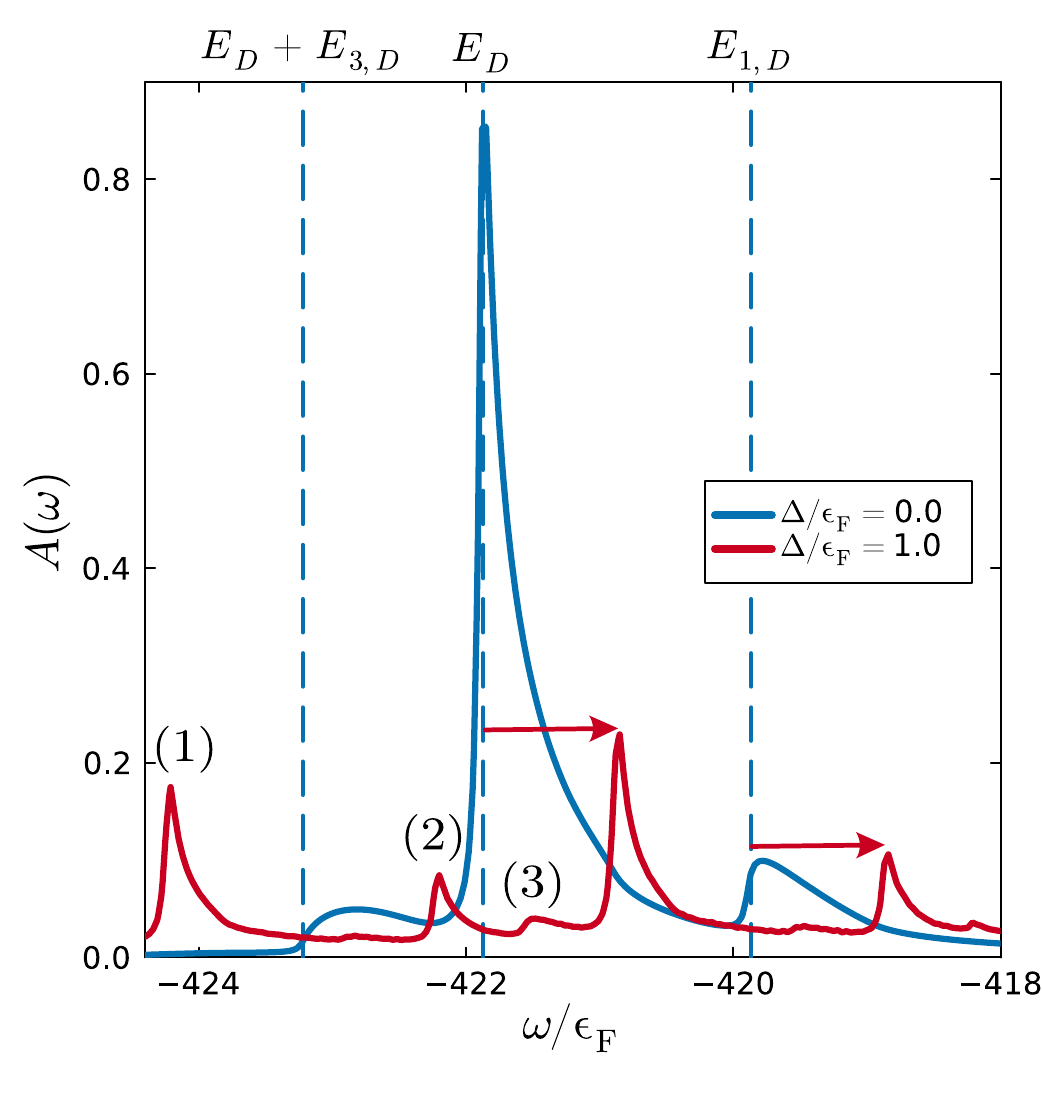}\hfill\\
\vspace{-.5em}
\caption{Vicinity of the dimer state of the deepest bound state with energy $E_D$; a magnified version of \Fig{fig:spectrum}(c) for exemplary gap strengths. Using the spectrum with $\Delta = 0$ for reference, we observe the appearance of multiple peaks for $\Delta>0$ corresponding to the occupation of the weakly bound state. The numbers indicate the energies of the secondary peaks: (1) the occupation of a tetramer at $E_D+\Delta+E_{3,T}^{(\Delta)}$, (2) a tetramer with $E_{1,D}+\Delta+E_{3,T}^{(\Delta)}$, and (3) a trimer with $E_{1,D}+\Delta+E^{(\Delta)}_{3,D}$. The read arrows point to the energies of the primary peaks without the occupation of the weakly bound states, i.e., $E_D+\Delta$ and $E_{1,D}+\Delta$.}
\label{fig:spec_dimer}
\end{figure}
Let us return to the discussion of the peak structure around the dimer and trimer peak corresponding to the deepest bound state (cf.\ \Fig{fig:spectrum}). Here, we characterize the secondary peaks that appear in the vicinity of the dimer state, as shown in \Fig{fig:spectrum}(c) or, for our purpose, magnified in \Fig{fig:spec_dimer}. For $\Delta=0$, we observe both dimer peaks for the scattering states $\ell=0,\, 1$ denoted by $E_D$ and $E_{1,D}$, respectively. These correspond to the occupation of a single bath fermion (after breaking a Cooper pair). Furthermore, a peak corresponding to the occupation of the deep bound state $E_D$ and the weak bound state $E_{3,D}$ is also apparent. For $\Delta>0$, doubly occupying the weak bound state (Cooper-pair capture) becomes more probable than single occupation, as seen by the spectral weights in~\Fig{fig:anomalous}, such that the highest secondary peaks will correspond to the occupation of the deepest bound state with a double occupation of the weakest one. Specifically, the configurations are:
\begin{itemize}
\item[$(1)$] Tetramer: $E_D+\Delta+E^{(\Delta)}_{3,T}$,
\item[$(2)$] Tetramer: $E_{1,D}+\Delta+E^{(\Delta)}_{3,T}$.
\end{itemize}
We can also identify a third peak corresponding to a single occupation of the weakest bound state with energy:
\begin{itemize}
\item[$(3)$] Trimer: $E_{1,D}+\Delta+E^{(\Delta)}_{3,D}$.
\end{itemize}

\subsection{Trimer}
\begin{figure}
\centering
\includegraphics[width=0.48\textwidth]{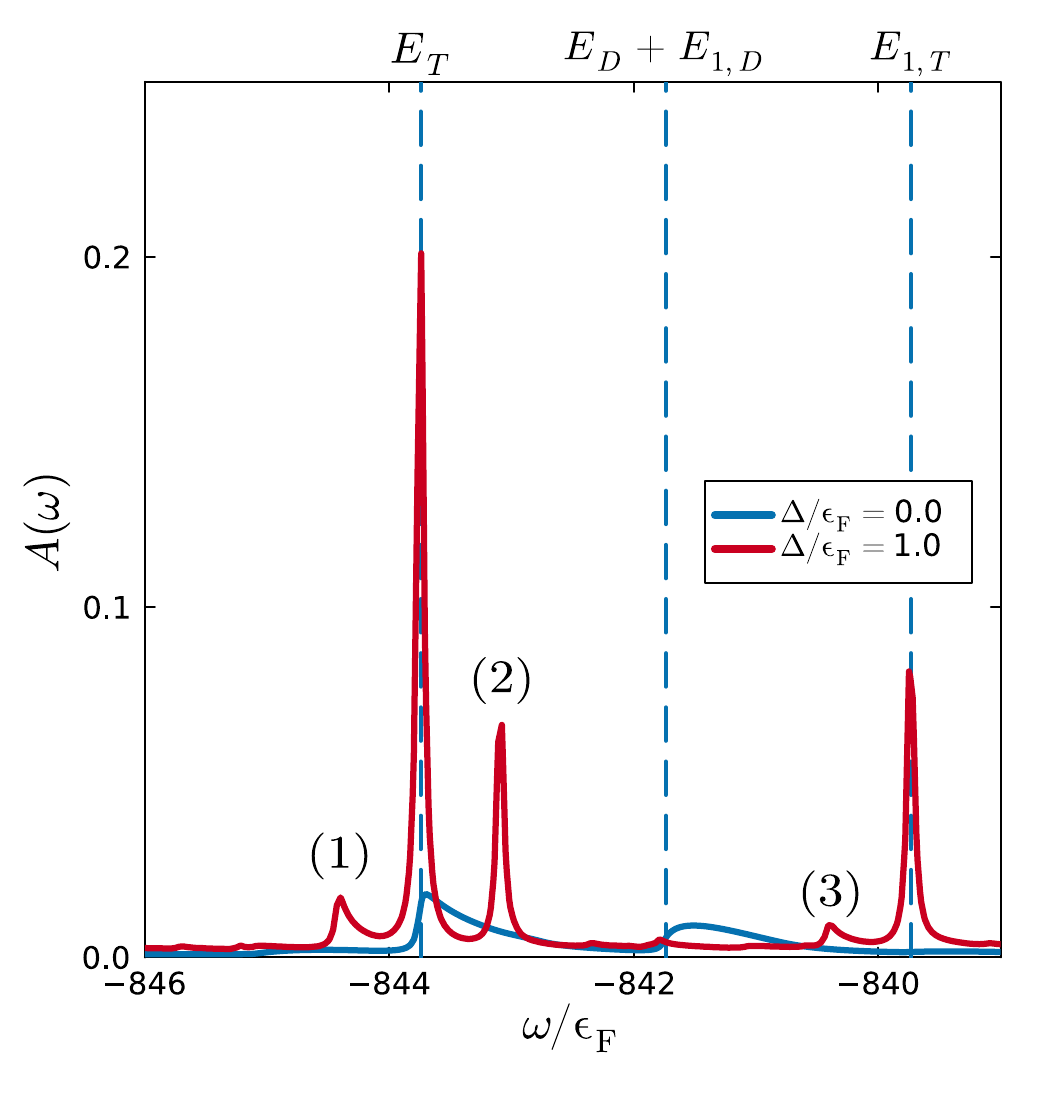}\hfill\\
\vspace{-.5em}
\caption{Vicinity of the trimer state of the deepest bound state with energy $E_T=2E_D$; a magnified version of \Fig{fig:spectrum}(b) for exemplary gap strengths. Using the $\Delta=0$ spectrum for reference, we observe the appearance of multiple peaks for $\Delta>0$ corresponding to the weak bound state occupation. The numbers indicate the energies of (1) the occupation of a tetramer with $E_T+E_{3,D}^{(\Delta)}$, (2) a pentamer with $E_{1,T}+E^{(\Delta)}_{3,T}$, and (3) a tetramer with $E_{1,T}+E^{(\Delta)}_{3,D}$.}
\label{fig:spec_trimer}
\end{figure}
Similarly to the dimer case, we can relate the peaks appearing in the trimer profile corresponding to the deepest bound state; see \Fig{fig:spec_trimer}, the magnified version of \Fig{fig:spectrum}(b). Again, we can identify the single or double occupation of the weakest bound state on top of the occupation of the trimer states detailed in the main text. Their specific energies are as follows:
\begin{itemize}
\item[$(1)$] Tetramer: $E_T+E^{(\Delta)}_{3,D}$,
\item[$(2)$] Pentamer: $E_{1,T}+E^{(\Delta)}_{3,T}$,
\item[$(3)$] Tetramer: $E_{1,T}+E^{(\Delta)}_{3,D}$.
\end{itemize}
Note that there are further spectral features visible with very small signal strength. These can also be identified analogously to the preceding discussion using weakly bound states. However, as they are so weak in strength, we defer from discussing these in further detail. 
Furthermore, there are other possible scenarios that exist, e.g., breaking a Cooper pair and trapping both particles, but these do not result in clear, distinct peaks in
our spectra due to their small many-body wave-function overlap.

\section{Quasiparticle weight}\label{sec:qw}
To obtain the quasiparticle weight, we need the long-time dynamics of the Ramsey signal. To this end, we can lower the time resolution since we only analyze the main emergent frequencies, i.e., bath and Bogoliubov branches, arising in the signal. To reduce the noise coming from the overlap of high energy bound states, we perform a Gaussian filter to the signal and then fit the resulting signal to Eq.~\eqref{eq:fit} from the main text. We can then compare the resulting power-law decay with the one obtained by means of the scattering phase shift. For a system with extension $R$ with vanishing boundary conditions, the scattering phase shift for each angular component follows as~\cite{griffiths_introduction_2018}
\begin{equation}
\delta_{\ell}(kR)=\arctan\left[\frac{j_{\ell}(kR)}{\eta_{\ell}(kR)}\right]\ ,
\end{equation}
where $j_{\ell}(\epsilon)$ and $\eta_{\ell}(\epsilon)$ are the spherical Bessel and Neumann functions, respectively.

\end{document}